\documentclass[twocolumn]{aastex631}

\usepackage[T1]{fontenc}

\begin{document}

\title{Reading between the rings: observed dust ring properties as probes of planet masses
}

\author[0009-0008-3992-1890]{Amena Faruqi}
\affiliation{Department of Physics, University of Warwick, Coventry CV4 7AL, UK}
\affiliation{Centre for Exoplanets and Habitability, University of Warwick, Coventry CV4 7AL, UK}

\author[0000-0003-3430-3889]{Jessica Speedie}
\altaffiliation{Heising-Simons Foundation 51 Pegasi b Fellow}
\affiliation{Department of Physics \& Astronomy, University of Victoria, Victoria, BC, V8P 5C2, Canada}
\affiliation{Department of Earth, Atmospheric, and Planetary Sciences, Massachusetts Institute of Technology, Cambridge, MA 02139, USA}

\author[0000-0002-7605-2961]{Ralph E. Pudritz}
\affiliation{Department of Physics and Astronomy, McMaster University\\
Hamilton, ON L8S 4K1, Canada}
\affiliation{Origins Institute, McMaster University, Hamilton, ON L8S 4K1, Canada}

\author[0000-0002-3984-9496]{Farzana Meru}
\affiliation{Department of Physics, University of Warwick, Coventry CV4 7AL, UK}
\affiliation{Centre for Exoplanets and Habitability, University of Warwick, Coventry CV4 7AL, UK}



\begin{abstract}
We hypothesise that dust rings in protoplanetary discs formed by an embedded planet should have properties that reflect the planet's mass. We use 2D hydrodynamical simulations of planet-disc interactions to investigate this, focusing on planets ranging $0.5-2.0\times$ the pebble-isolation mass, for three different aspect ratios.  We find the ring's dust mass, peak location, and width to correlate with planet mass. We confirm a positive linear relationship between a planet's Hill radius and the location of a ring's density peak and demonstrate how this relationship can be used to constrain planet masses in observed systems by applying it to PDS 70. The dust ring width and mass change with planet mass for planet masses up to the pebble-isolation mass, beyond which they become constant. The steepness of the gas pressure radial profile is asymmetric, with the direction of the asymmetry being determined by whether the planet mass is above or below the pebble-isolation mass. We therefore propose a new way to define the pebble-isolation mass: the minimum planet mass which perturbs the gas enough for the pressure gradient interior to the pressure maximum to exceed the pressure gradient exterior to it. We discuss how our findings could be used to constrain or estimate planet masses from gas or dust observations of discs with measurable substructures and apply our results to 5 discs in the exoALMA sample to estimate planet masses and constrain disc aspect ratios. We also discuss how the potential for planetesimal formation in a ring varies with planet mass. \\

\end{abstract}



\section{Introduction} \label{sec:intro}

Since they were first imaged in the 1990s, protoplanetary discs have been the subject of great interest as the putative birthplace of planetary systems like our own. More recent observations, such as those performed by the Atacama Large Millimeter/submillimeter Array (ALMA) have demonstrated that protoplanetary discs display a significant diversity of sizes, shapes, alignments and substructures \citep[e.g.][]{alma2015, huang2018, dullemond2018, exoalma2025}. Of these, the prevalence of gaps and pileups (rings) in the distribution of the solid matter of discs has sparked interest in the hypothesis that these may indicate the presence of planets embedded within these discs \citep[e.g.][]{goldreich1980, lin1993, paardekooper2004}.\\ 

It is certainly possible that these features could be formed via other means that do not require a planet. For instance, they could be formed as a result of rapid dust growth due to the condensation of volatiles \citep{zhang2015}, material flows at the dead-zone edge \citep{flock2015}, as a consequence of gap clearing by MHD disk winds \citep{Riols_Lesur2019}, or through a self-induced dust pileup \citep{gonzalez2015b,gonzalez2017}, to name a few. However, if even some fraction of these observed rings are caused by fully-formed planets, it is important to explore how observations and measurements of such rings can inform our understanding of these nascent planets and the planetary systems they may go on to comprise. Detailed modeling can also expand our understanding of the theoretical underpinnings of how an embedded planet influences the material in its disc and how that, in turn, can influence the disc's ability to form further planetesimals. \\

A concrete example of a planet-induced ring is in the observed PDS 70 system \citep{Keppler2018, Haffert2019}.  \citet{Doi2024} measured the lower limit of total dust mass contained in PDS 70's ring using ALMA Band 3 data and found it to be around $28 M_\oplus$, proving that rings can contain sufficient mass to form planets. In another example, the first measurements of ring widths and masses using the DSHARP data \citep{dullemond2018} revealed at least two critical features about rings: that they are narrow, indicating low levels of turbulence ($\alpha \le 10^{-4}$), and that they can trap tens of Earth masses of dust ($\sim 40 M_\oplus$).  How $\sim 20 - 40 M_\oplus$ of material, enough to form multiple solid planetary cores via collisional growth, can pile up in these rings is important to investigate.  \\
 
 In this regard, one critical detail about rings in the context of planet formation is that the collision rate between dust grains increases greatly within them. If the dust-to-gas ratio reaches a critical value, planetesimal formation can be triggered. Specifically, the heightened dust-to-gas ratio in the ring can trigger the streaming instability \citep{youdin2005}, whereby large concentrations of dust are able to shield one another from the gas drag, preventing them from drifting inwards faster than they can grow, and allowing them to eventually undergo gravitational collapse to form solid clumps, which can grow further into planetesimals.  In general, the feasibility of these processes depends strongly on properties such as the mass of the initial planet (and its ability to enhance the dust-to-gas ratio) as well as key properties of the protoplanetary discs such as their aspect ratio, turbulence level, and the effective Stokes number of the drifting dust. Therefore studying this link is integral to understanding the planetary systems we observe today.\\ 

Measurements of disc substructures have been possible for several years now, with properties such as gap and ring  widths, gap depths, and semi-major axes being constrained with high precision and accuracy for multiple well-resolved discs. For instance studies of gap properties \ \citet{kanagawa2016} found an empirical relationship between the mass of a planet and the width of the gas gap it produces, as well as demonstrating how this could be applied to the gaps in the protoplanetary disk in HL Tau to infer the masses of planets that could be present. Building on this, \citet{zhang2018} found an empirical relationship between planet mass and gas gap width and depth, with which they found good agreement  between models and observations of the AS 209 system.\\ 

That said, ring properties in their own right can also provide important diagnostic information about planets that may have formed them, but have received less attention.  Thus, \citet{pinilla2012} found a relationship between a high mass planet's Hill radius ($R_{H}$) and the location of a dust ring, which \citet{rosotti2016minimum} later confirmed and found to be linear within a range of planet masses spanning several Earth masses. \citet{lodato2019} found that their measurements of the dust gap outer edges of CI Tau and MWC 480 were also consistent with these past results. \\

Overall, unraveling the link between these measurements of structural properties and the inferred properties of the planets that may have caused them has been limited mainly to how planet masses relate to gap properties. The properties of dust rings and how they scale with planet masses that induce them, the physical properties of the disk (turbulence level, scale height), as well as Stokes numbers of the dust is equally important.  One of the central insights of planet formation theory is that dust rings build up in pressure bumps induced by the planet.  If planets grow by pebble accretion, theory \citep{bitsch2018, lambrechts2014} predicts that pebble flows through the rings are shut off when planets achieve the pebble isolation mass.  We note that this may limit or control the mass of solid material in the rings.  It is the purpose of this paper to explore the connections between rings and planets in much greater depth.  \\  

In this paper, we run 2D hydrodynamical models of a protoplanetary disc with an embedded planet to quantify how measurable properties of dust rings relate to the mass of the planet, for planets above and below the pebble-isolation mass. We consider different disc aspect ratios to relate our findings to the theory of pebble-isolation mass. Section \ref{sec:theory} outlines the concept of a pebble-isolation mass and how it relates to ring formation. Section \ref{sec:methods} describes the numerical methods employed to model the discs, as well as the chosen model parameters and simulation setup. The findings from these simulations and how they can be applied to observations are discussed in Section \ref{sec:results}. We discuss the implications of our results and demonstrate how they can be applied to observed systems in Section \ref{sec:discussion}, then summarize our results and present our main conclusions in Section \ref{sec:conclusion}.\\

\section{Pebble-Isolation Theory}
\label{sec:theory}

Ring and gap formation by planetary perturbations requires that a sufficiently massive planet is capable of exchanging angular momentum with gaseous material in the region of its orbit. Some of this material is expelled, opening a partial gap in the gas density distribution. This, in turn, produces a local pressure maximum exterior to the planet's orbit, since the gas pressure traces the gas surface density profile as
\begin{equation}
    \label{eq:pressure}
    P = \frac{c_{s}^{2}\Sigma_{g}}{h}
\end{equation}
where $c_{s}$ is the sound speed in the gas, $\Sigma_{g}$ is the gas surface density, and $h$ is the disc scale height.\\
 
This perturbation to the gas component exerts forces on the dust that oppose its radial inflow, causing it to slow and collect at the pressure maximum. The radial velocity of dust in these rings is determined by the magnitude of the pressure gradients \citep{weidenschilling1977a}:
\begin{equation}
    \label{eq:vr}
    v_{r} = \frac{c_{s}}{\rm St + \rm St^{-1}} \frac{h}{r} \frac{\partial \ln P}{\partial \ln r}
\end{equation}
where St is the Stokes number of the dust, and $r$ is the radial location in the disc. On either side of a pressure maximum, $\partial P/\partial r$ will have opposite signs, creating converging flows of dust. If the pressure gradients are sufficient for slowing or stopping the inwards radial drift of dust, dust grains pile up in a ``dust trap'', seen in observations as a concentrated ring of dust.  \\

One of the key concepts in the pebble accretion picture of planet formation is that there is a minimum planet mass capable of trapping pebble-sized dust (Stokes numbers of $\sim 0.1-1$) into a ring - known as the ``pebble-isolation mass'' \citep{morbidelli2012, lambrechts2014}. Planets of this mass and greater can perturb the gas disc enough to prevent the radial inflow of pebble-sized dust from the outer disc, leading to observable consequences. For instance, an efficient dust trap may produce distinct reservoirs of material on either side of the planet that could explain the dichotomy of terrestrial and giant planets seen in our own Solar System \citep{morbidelli2015, Alibert2018}. ALMA observations, such as those highlighted by the DSHARP survey \citep{dullemond2018}, showed bright dust rings present in a number of discs, which could also demonstrate this mechanism at play. \\

A number of different prescriptions have been presented for the pebble-isolation mass. \citet{lambrechts2014} first proposed a scaling relation that linked the pebble-isolation mass to the disc's aspect ratio. \citet{bitsch2018} later conducted a detail parameter study, deriving an expression for the pebble-isolation mass that varies with the disc's $\alpha$-viscosity and initial pressure gradient, and accounts for the different behaviour of dust grains with different Stokes numbers. They also find that a coefficient of $20 M_\oplus$, rather than the $25 M_\oplus$ used by \citet{lambrechts2014}, fits their models better. \citet{ataiee2018} carried out similar work and produced a prescription for the pebble-isolation mass that is parameterised by the same quantities but also accounts for the stellar mass. The work of \citet{bitsch2018} and \citet{ataiee2018} is largely in agreement, with the formula developed by \citet{ataiee2018} providing slightly lower but comparable estimates for the pebble-isolation mass. In this work, we follow the prescription given by \citet{bitsch2018}, since it takes into account a broad range of parameters, is well-tested, and is consistent with the prior results of \citet{lambrechts2014}.\\

It should be stressed that the pebble-isolation mass is often treated as a distinct boundary, whereby planets with masses below the pebble-isolation mass do not form a ring and planets with masses above the pebble-isolation mass do. However, it unlikely that the boundary defined by the pebble-isolation mass is as sharp as it is often treated and a detailed study into how properties of dust rings change as a planet's mass is gradually increased up to and beyond the pebble-isolation mass is essential for understanding this better.  \\

\section{Numerical Methods} \label{sec:methods}
\subsection{Overview}
We prescribe the pebble-isolation mass in this  paper using the detailed, semi-analytical prescription derived by \citet{bitsch2018} and based on numerical experiments for viscous disks:

\begin{equation}
\label{eq:miso}
    M_{\rm iso} = \bigg(25 + \frac{\alpha/2 \rm St}{0.00476} \bigg)f_{\rm fit}  M_\oplus
\end{equation}\\
where
\begin{equation}
\label{eq:ffit}
    f_{\rm fit} = \Bigg[ \frac{h_{p}/r_{p}}{0.05} \Bigg]^{3} \Bigg[ 0.34 \bigg( \frac{\log \alpha_{3}}{\log \alpha} \bigg)^{4} + 0.66 \Bigg] \Bigg[ 1 - \frac{\frac{\partial \ln P}{\partial \ln r}+2.5}{6} \Bigg]
\end{equation}
with $h_{p}/r_{p}$ is the disc aspect ratio at the planet's radial location $r_{p}$, $\alpha_{3} = 0.001$, $\alpha$ is the Shakura-Sunyaev $\alpha$-viscosity parameter \citep{shakura1973}, and $\partial \ln P/ \partial \ln r$ is the disc's unperturbed gas pressure gradient.\\

It is necessary to test a range of values for the disc aspect ratio to ensure that any inferred link between the pebble-isolation mass and the properties of dust substructures holds for all aspect ratios. On this basis, we select three aspect ratios at the planet's location in our simulations: the canonical 0.05, and two additional values, 0.06 and 0.07. We then use Equation \ref{eq:miso} to compute the pebble-isolation mass for each of these aspect ratios, for a Stokes number of 0.1 and an $\alpha$-viscosity of $10^{-4}$. We define a range of planet masses to model for each aspect ratio such that we cover masses above and below the pebble-isolation mass. The calculated masses are shown in Table \ref{tab:masses}.\\

\begin{table}[]
\centering
    \begin{tabular}{|c||cccccc|}
    \hline
    $M/M_{\rm iso}$  & 0.50 & 0.75 & 1.00 & 1.25 & 1.50 & 2.00 \\ \hline \hline
    $h_{p}/r_{p} = 0.05$ & 10            & 15            & 20            & 25            & 30            & 40            \\ \hline
    $h_{p}/r_{p} = 0.06$ & 20            & 30            & 40            & 50            & 60            & 80            \\ \hline
    $h_{p}/r_{p} = 0.07$ & 30            & 45            & 60            & 75            & 90            & 120            \\ \hline
    \end{tabular}
    \caption{Chosen planet masses expressed as a fraction of the pebble-isolation mass for each aspect ratio and in units of Earth masses. The stellar mass across all models is $1 M_\odot$.}
    \label{tab:masses}
\end{table}

\subsection{Simulation setup}  \label{sec:simsetup}
We run a series of 2D ($r + \phi$) simulations of gas and 5 dust species using \textsc{FARGO3D} \citep{masset2000fargo, llambay2016fargo}. In all models, the gas and dust are evolved at the same time, with the dust being modeled as a pressureless fluid subject to gas drag. Dust feedback has been neglected in these simulations. In total, 18 models are run (see Table \ref{tab:masses}) to 1500 planet orbits, by which point we find that the dust substructures have reached an approximate steady-state. All models are scale-free, with the planet on a fixed orbit at $r_{p}=1$ on a grid spanning 0.2 to 3$r_{p}$. To ensure that the planet-induced perturbations are sufficiently resolved, we run all models with a minimum grid resolution of 4-6 cells per planet Hill radius and 9-13 cells per disc scale height at the planet's location. This corresponds to cylindrical grid dimensions of 504 radial cells against 1170 azimuthal cells, with the grid cells being logarithmically spaced in the radial direction to maintain an approximately square cell geometry. \\

The initial surface density of the gas component of the disc is defined as
\begin{equation}
    \Sigma_{g} = \Sigma_{0} \bigg (\frac{r}{r_{0}} \bigg )^{-1}
\end{equation}
where $\Sigma_{0}$ is the dimensionless surface density at $r=r_{0}=1$ with the value of $1 \times 10^{-3}$. Since these models are scale-free, this value is arbitrary and can be scaled to physical units via any chosen scaling relation. For example, by treating the central star mass as $1 M_\odot$ and rescaling the planet's location to a value in physical units of $50$ AU, $\Sigma_{0} = 1 \times 10^{-3}$ would correspond to a surface density of $3555 g/cm^{2}$ at 1 AU and a disc mass of $0.018 M_\odot$. \\

The 5 chosen dust species have fixed Stokes numbers that are logarithmically spaced from 0.002 to 0.2 i.e. 0.002, 0.0063, 0.02, 0.063, and 0.2. The initial dust surface density is uniform across all dust species: it is equal to the initial gas surface density multiplied by the dust-to-gas mass ratio, which is 0.01. This means that the total dust mass in the simulations is, in actuality, 5\% of the total gas mass, rather than the canonical 1\% dust-to-gas ratio taken from the ISM \citep{bohlin1978ism} and applied to protoplanetary discs. However, since the different dust species do not interact with one another or with the gas, this has no impact on the results and can be accounted for during analysis, via methods outlined in Section \ref{sec:dmass_calc}. The $\alpha$-viscosity for all models is $10^{-4}$, in accordance with the growing consensus that $\alpha$-viscosities of observed discs are most likely $\leq 10^{-3}$ \citep[e.g.][]{lesur2023}. Antisymmetric boundary conditions are applied to the gas and dust radial velocities at both the inner and outer boundaries of the disc to simulate realistic inflows and outflows at these locations.  \\

\subsection{Ring identification}  \label{sec:ringfitting}

\begin{figure}
    \centering
    \includegraphics[width=0.95\linewidth]{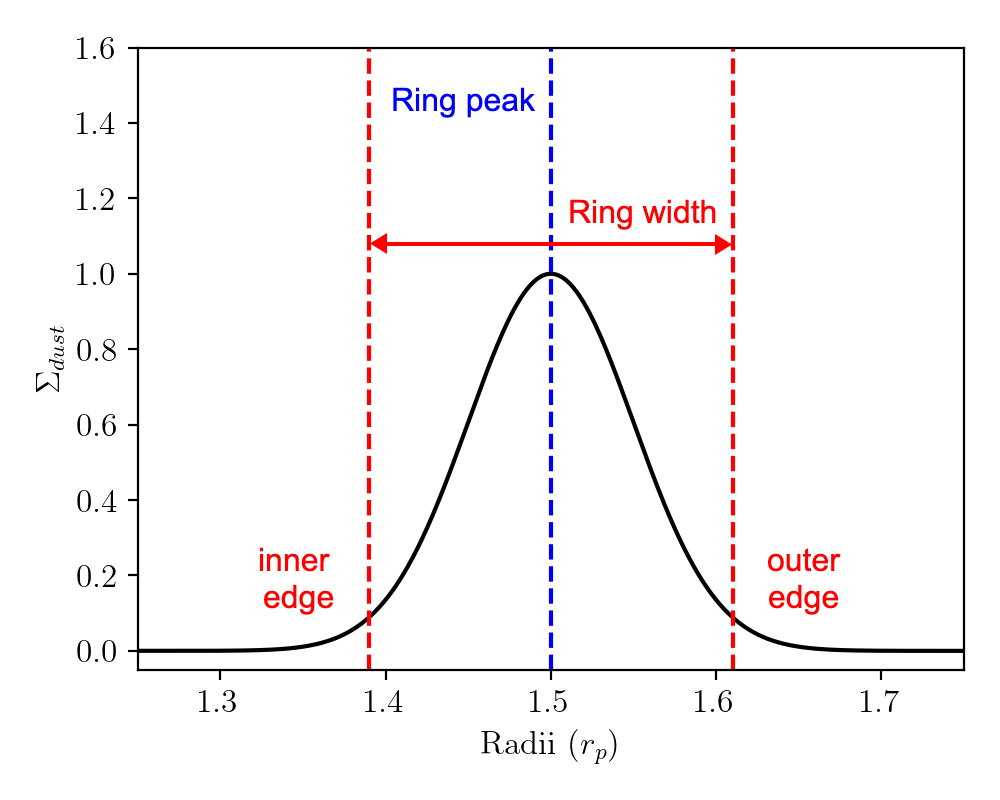}
    \caption{Schematic showing the different ring properties being measured in simulations and how they are defined.}
    \label{fig:ring_schematic}
\end{figure}
To determine properties of rings, we use three measurable quantities: the azimuthally-averaged radial location of the peak dust density within the ring, and the azimuthally-averaged radial location of the inner and outer edges of the ring (see Figure \ref{fig:ring_schematic}). The distance between the ring outer edge and inner edge is taken to be the ring width. Due to the asymmetric density profiles in some of these rings (as shown in Figure  \ref{fig:1dsigmas}), the ring peak does not necessarily sit at the exact midpoint between the inner and outer ring edges so needs to be identified independently. \\

To identify the ring density peak, we  restrict our search to a radial range that we expect to contain the ring, to avoid misidentifying global maxima. Since it has been shown that the outer edge of a dust gap scales with the Hill radius of the planet \citep[e.g.][]{rosotti2016minimum}, we choose the range to scale with the planet's Hill radius. We expect the ring to fall within 3 and 20 Hill radii of the planet's location -- the outer bound is left intentionally wide to ensure we capture even disperse, shallow rings, while the same cannot be done for the inner bound without including the dust gap and a potential density maximum interior to it in the analysis. We normalise the dust surface density profile to account for the unperturbed shape of the density profile, leaving us with a profile that is flat everywhere other than where the dust has been perturbed by the planet (see Figure \ref{fig:1dsigmas} for examples of the normalised profiles used). Then, the density peak is identified as the local maximum of the dust surface density (within the bounded region). \\

One common method for identifying the edges of a dust ring is to apply a Gaussian or Lorentzian fit to its shape and take the edges to be some integer number of standard deviations from the mean. However, as noted previously, the density profiles of rings are not necessarily symmetric about the density peak, as has also been seen in previous studies that sought to quantify ring properties  \citep[e.g.][]{dullemond2018}. Instead, we define our own criteria for identifying the ring edges. Starting from the ring peak, we move radially inwards or outwards from the peak until one of two criteria are met: either the value of the surface density drops to 10\% of its peak value, or the magnitude of the surface density gradient drops below an arbitrarily small value, which was chosen here to be $0.5$. This second condition was necessary to account for instances where the dust surface density at the outer edge of a shallow ring tapers off very gradually, and therefore does not reach the threshold value of 10\% of the peak density within the bounded region. Changing these threshold values does yield different values for the measured ring widths but the scaling with planet mass we find in Section \ref{sec:results} still holds. An example of the ring edges identified via this method can be seen in Appendix \ref{app1}.  \\

\subsection{Dust mass calibration}
\label{sec:dmass_calc}
Prior to computing the dust mass, we must first apply a weighting to the mass in each dust bin to ensure that the total dust-to-gas ratio of all dust is 0.01, such that
\begin{equation}
    \Sigma_{s} = \sum^{N_{dust}}_{i=1} W_{i} \Sigma_{d,i} 
\end{equation}
and 
\begin{equation}
    \sum^{N_{dust}}_{i=1} W_{i} = 1
\end{equation}
where $\Sigma_{s}$ is the total density of solid matter, $N_{dust}$ is the number of dust species, $W_{i}$ is the size-based weighting applied to the $i^{th}$ dust species, and $\Sigma_{d,i}$ is the surface density of the $i^{th}$ dust species, which is initially uniform across all dust sizes. \\

The values of $W_{i}$ are equal to the proportion of the total dust mass we have in each size bin, which is specified by the dust size distribution given in \citet{Mathis1977} (hereafter referred to as the ``MRN'' distribution). In order to calculate the values of $W_{i}$ for each dust species, we begin by integrating the MRN distribution to obtain the number density of grains of physical size $a$:
\begin{equation}
    n(a) \propto a^{-2.5}
\end{equation}

By multiplying this by the mass of a single grain of size $a$ (which is proportional to $a^{3}$), we find that the total mass contained in the $i^{th}$ size bin is proportional to $a_{i}^{1/2}$. We can therefore express $W_{i}$ as the total mass in the $i^{th}$ size bin divided by the sum of the dust mass across all size bins:
\begin{equation}
    W_{i} = \frac{a_{i}^{1/2}}{\sum^{N_{dust}}_{j=1} a_{i}^{1/2}} = \frac{\rm St_{i}^{1/2}}{\sum^{N_{dust}}_{j=1} \rm St_{i}^{1/2}}
\end{equation}

The dust mass contained within the rings for a particular Stokes number can then be calculated by integrating the dust surface density profile between the ring inner and outer edge and multiplying by the resultant mass by the weighting, $W_{i}$, for that Stokes number. The total dust mass across all Stokes numbers should then always be equal to 1\% of the total gas mass in the disc, although we note that the open inner boundary and antisymmetric outer boundary used in our simulations do permit material to leave the disc via the inner edge and to be replenished via the outer edge, changing the total dust-to-gas ratio over the course of the simulation. \\

\section{Results} \label{sec:results}
In the following sections, we outline the relationships we find between a planet's mass and the width and location of the ring it produces. We detail how these relationships may be used to estimate or constrain planet masses from observations of discs, and apply our methods to the PDS 70 system and several discs from the exoALMA survey to demonstrate this. Finally, we examine how the solid mass in the rings relates to planet mass and discuss what this tells us about the prospect of sequential planet formation and how this may be reflected in the architectures of observed planetary systems.\\

\begin{figure*}
    \centering
    \includegraphics[width=1\linewidth]{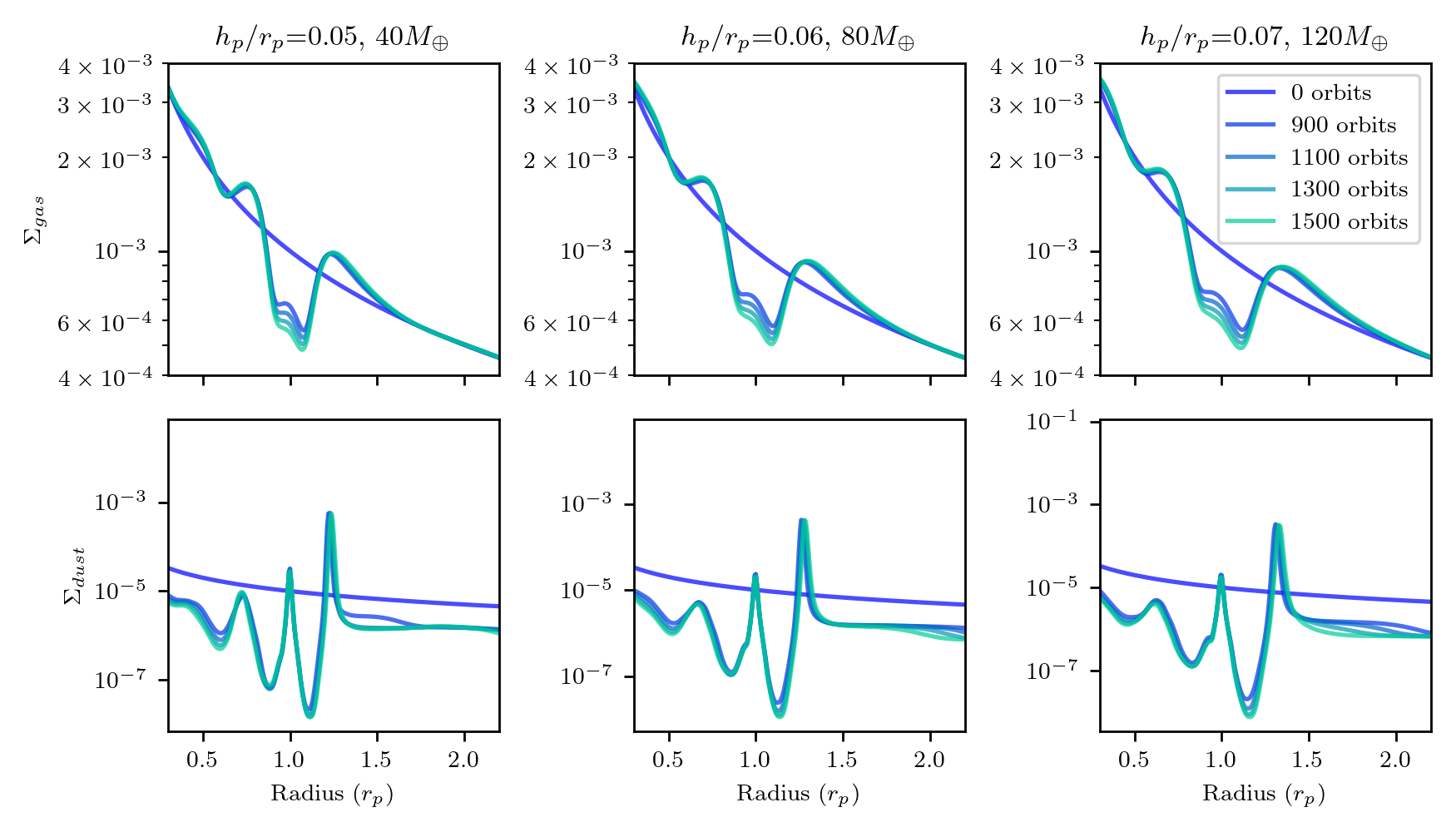}
    \caption{Time evolution of gas and dust surface densities of the highest planet mass model for each aspect ratio.}
    \label{fig:timeevo}
\end{figure*}

Before calculating the ring properties defined in the previous section, we first confirm that the runtime of the simulations, 1500 planet orbits, is sufficient for the rings to have reached a steady state. We plot the time evolution of the gas and dust surface densities of the highest planet mass models for each of the three aspect ratios in Figure \ref{fig:timeevo} (40, 80, and 120 $M_\oplus$ for $h_{p}/r_{p}$ of 0.05, 0.06, 0.07,respectively). We choose to only plot these for the last 600 orbits to specifically show how much the dust and gas varies towards the end of the simulations. We find that the ring surface mass density is unchanging and the dust rings have reached an approximately steady height and width by 1500 orbits and conclude that this runtime is sufficient for our analysis. \\

\subsection{Ring widths}
\label{sec:results:rwidth}


\begin{figure}
\centering
\includegraphics[width=\columnwidth]{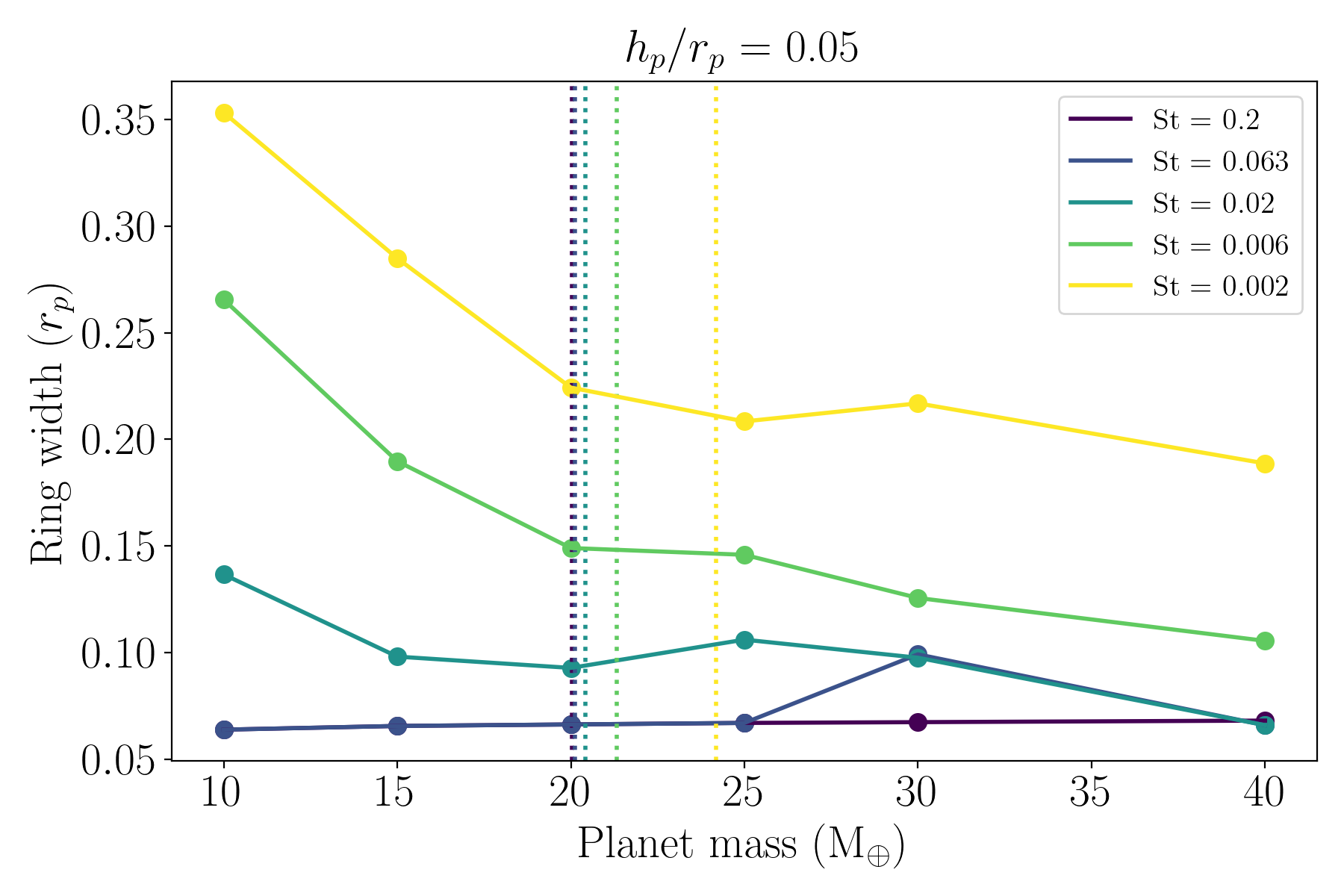}\par\vspace{1ex}
\includegraphics[width=\columnwidth]{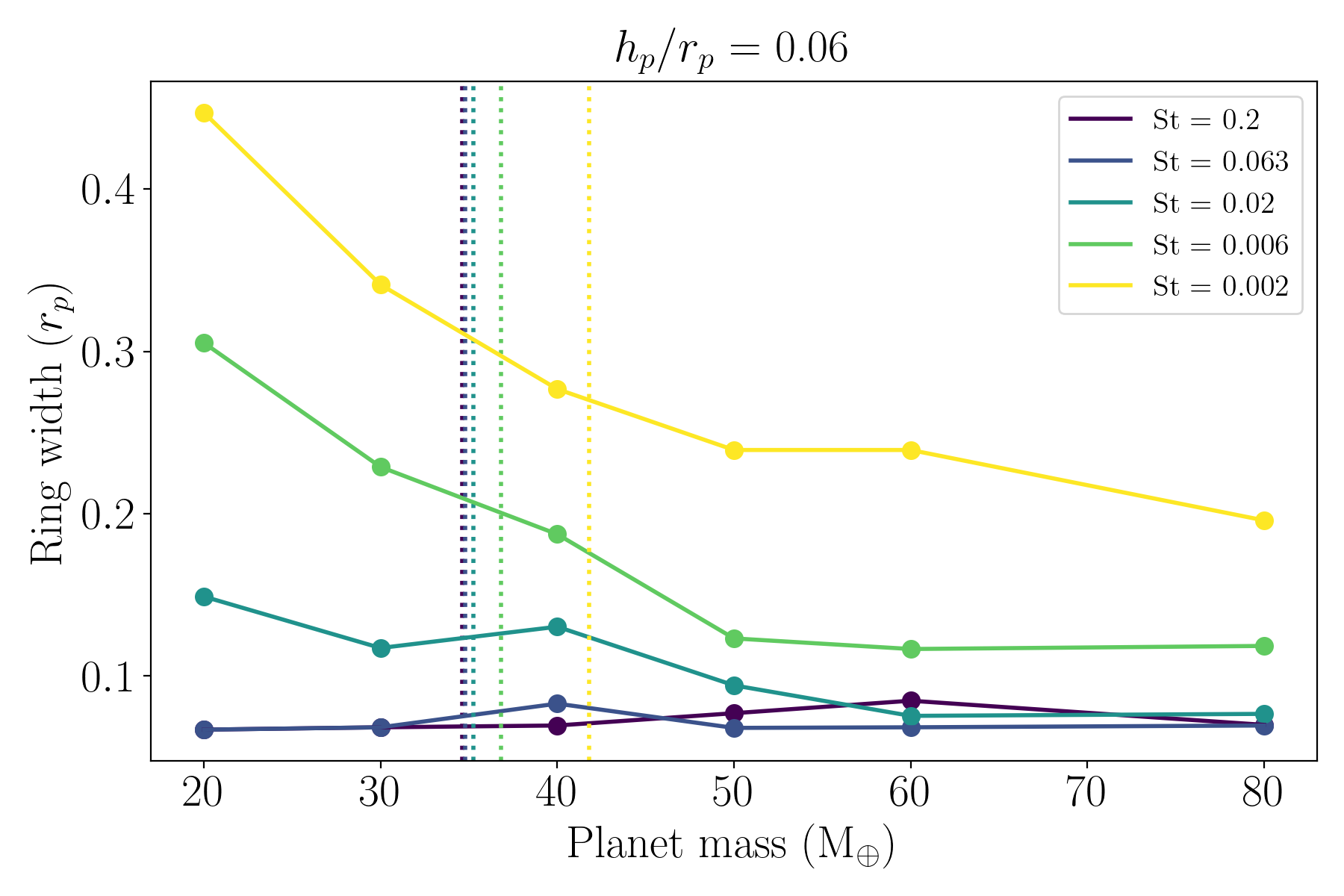}\par\vspace{1ex}
\includegraphics[width=\columnwidth]{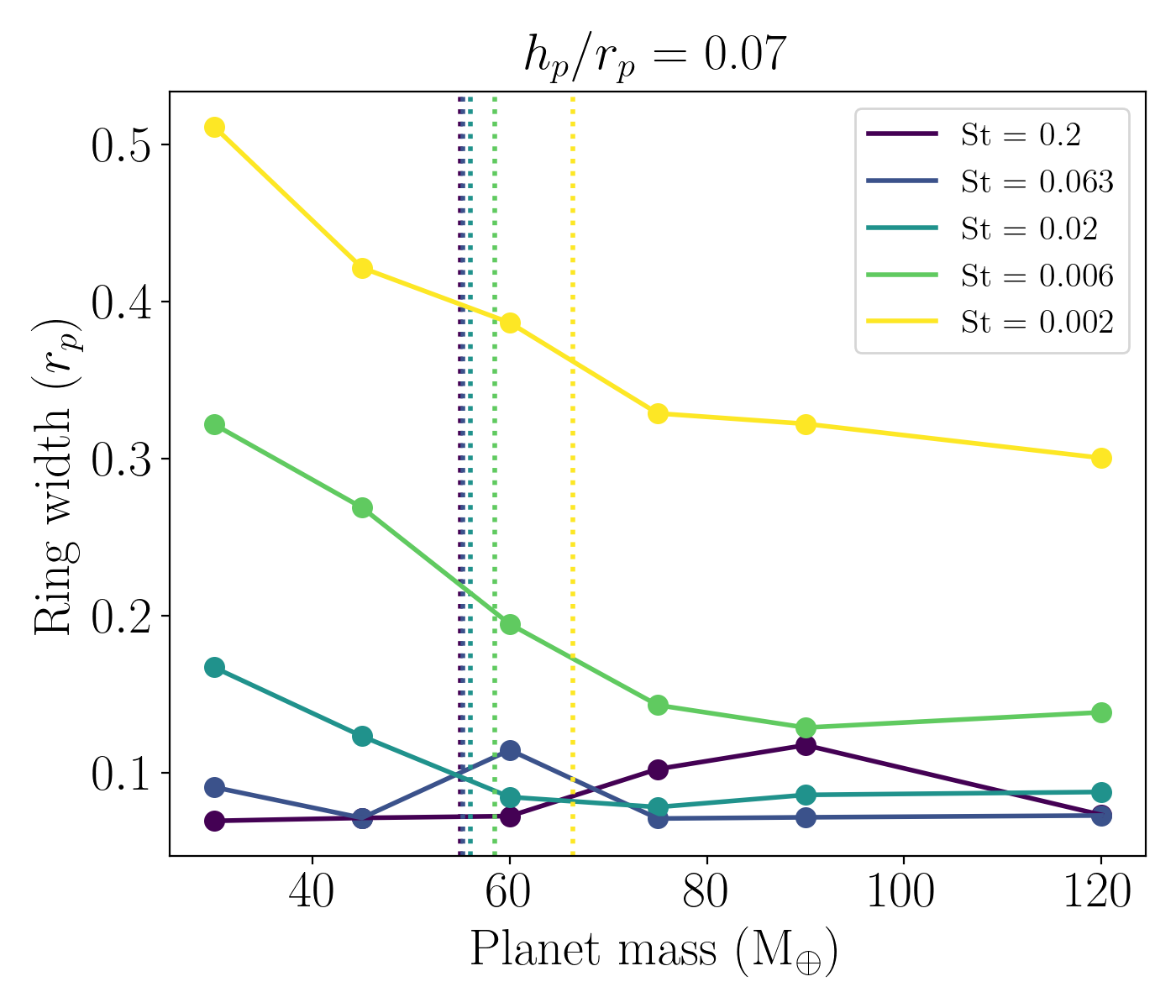}
\caption{Dust ring widths against planet masses for all three aspect ratios.
Vertical dotted lines indicate the pebble-isolation mass for each Stokes value,
calculated according to the prescription by \citet{bitsch2018}.
The ring widths for the two largest Stokes numbers are constant with increasing
planet mass, so overlap. Ring width decreases with increasing planet mass for
planet masses below the pebble-isolation mass, beyond which it is approximately
constant.}
\label{fig:rwidths}
\end{figure}

Figure \ref{fig:rwidths} shows the widths of the dust rings against planet masses for discs with aspect ratios of $h_{p}/r_{p}=0.05$, 0.06, and 0.07, respectively. We find that the ring width decreases with planet mass for planet masses up to the pebble-isolation mass, beyond which the ring width plateaus. This is consistently the case across all aspect ratios, for which the pebble-isolation mass differs, due to the steep dependence on the aspect ratio (see Equation \ref{eq:miso}). We note small deviations from this trend, such as in the case of the St=0.2 dust grains in the $h_{p}/r_{p}=0.07$ disc, where two largest planet masses an anomalous increase in the ring width. In all cases, these are caused the formation of ``shoulders'' or bumps on the interior or exterior edge of the ring, which leads to a perceived widening of the ring (see Figure \ref{fig:1dsigmas}).\\

We note that this relationship between the ring width and planet mass is most obvious when looking at small dust grains (St=0.002) and becomes less pronounced as the Stokes number increases, with St=0.2 grains not showing any variation in the ring width with planet mass at all. This is consistent with the fact that the pebble-isolation mass is lower for higher Stokes number grains, meaning even the lowest planet mass modeled for each aspect ratio is capable of trapping the largest dust grains in a narrow ring. \\

\subsection{Ring peak locations}
\label{sec:results_rpeak}


\begin{figure}
    \centering
    \includegraphics[width=1\linewidth]{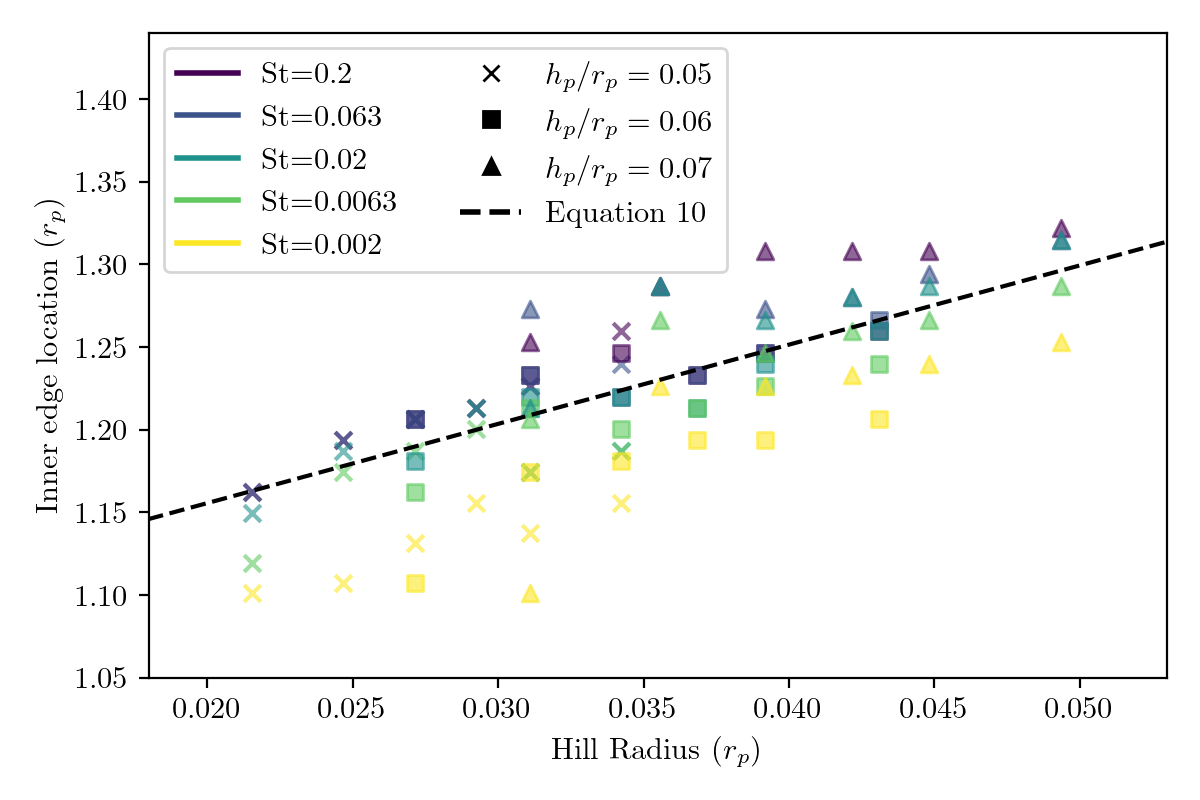}
    \caption{Locations of the dust ring inner edges against the planets' Hill radii for all three aspect ratios. The dashed black line indicates the linear fit applied to the data points, stated in Equation \ref{eq:redge}.}
    \label{fig:redges}
\end{figure}

\begin{figure}
    \centering
    \includegraphics[width=1\linewidth]{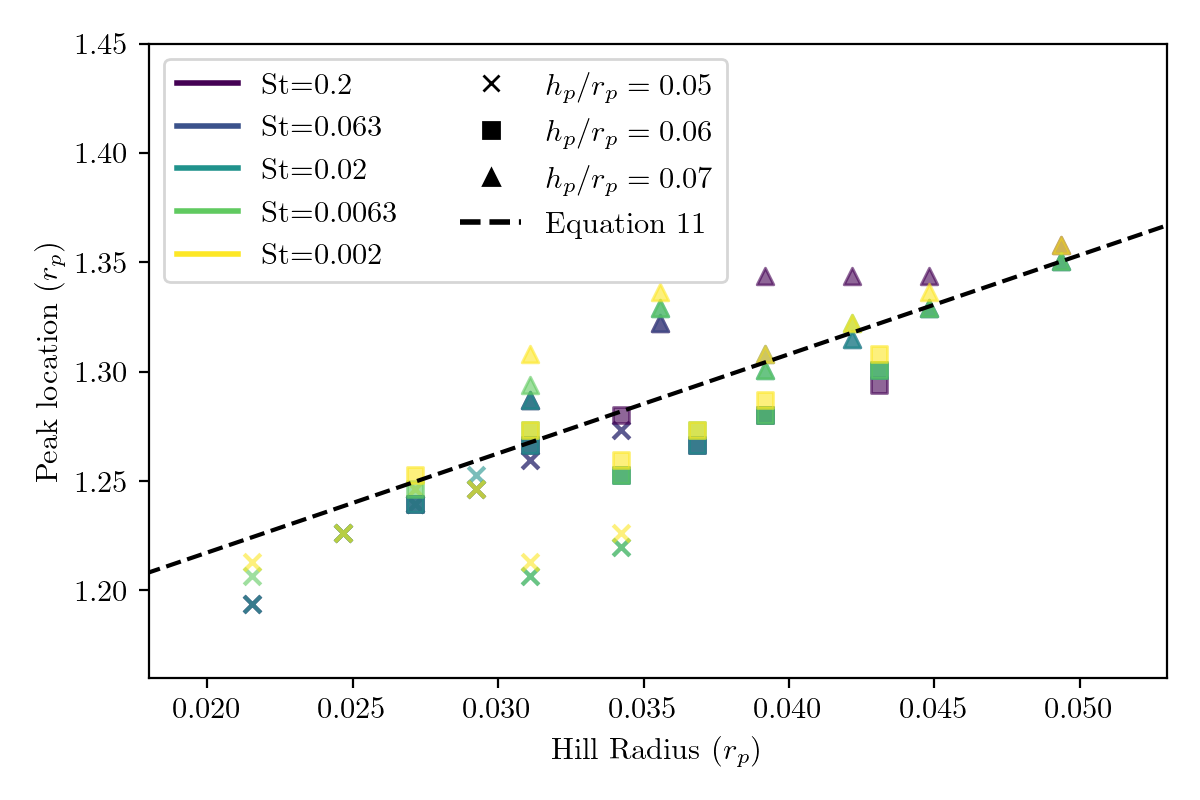}
    \caption{Locations of the dust ring peaks against the planets' Hill radii for all three aspect ratios. The dashed black line indicates the linear fit applied to the data points, stated in Equation \ref{eq:rpeak0}. }
    \label{fig:rpeaks}
\end{figure}

Figure \ref{fig:redges} shows the locations of the dust ring inner edges against the planets' Hill radii for each of the three aspect ratios modeled. In past work, \citet{rosotti2016minimum} found a linear relationship between a planet's Hill radius and the outer edge of the dust gap, which \citet{lodato2019} found to be consistent with their measurements of two observed systems. We confirm that the radial location of the ring's inner edge (equivalent to the outer edge of the dust gap) scales linearly with the planet's Hill radius and find the best fit to our data to be:
\begin{equation}
    \label{eq:redge}
    r_{edge} = (1.0 \pm 0.02)r_{p} + (5.4 \pm 0.5) R_{H}
\end{equation}
where $r_{edge}$ is the location of the outer edge of the observed dust gap. \\

Like \citet{rosotti2016minimum}, we find that this proportionality is independent of the disc's aspect ratio as it holds for all of our models. However, we find that the relationship between the planet's Hill radius and the dust inner edge is offset based on the Stokes number of the dust. We find that smaller dust ($\rm St \lesssim 10^{-3}$) collects at a ring inner edge that is closer to the planet, while larger dust ($\rm St \gtrsim 10^{-3}$) settles at an inner edge that is further away. Of our chosen Stokes values, St=0.0063 and St=0.02 most closely match the scaling relation provided by \citet{lodato2019}. Taking the ring to be located, on average, at approximately $r=1.3$, this corresponds to a physical grain size of $\sim 1.3-4.4$mm. The data used by \citet{lodato2019} comes from ALMA Band 6 observations, which have a wavelength of 1.3mm, so this is consistent with our findings.\\

As the planet mass increases and the inner edge moves outward, the ring peak is pushed outwards by approximately the same extent (shown by the linear trend in the data in both Figures \ref{fig:redges} and \ref{fig:rpeaks}, which have approximately the same slope). However, the ring outer edge location moves inwards relative to the ring peak with increasing planet mass, leading to the narrowing of the ring. This can be seen in the dust surface density profiles in Appendix \ref{app1}. This is consistent with our findings that the ring width decreases with increasing planet mass. When taking into consideration that the ring inner edge is closer in for smaller dust, this also explains why the measured ring width is highest for the smallest dust grains. \\

In Figure \ref{fig:rpeaks}, we plot the peak of the ring density and find that this is both proportional to the Hill radius and independent of Stokes number, for the range of Stokes values modelled (corresponding to physical sizes ranging from sub-mm to mm sized grains). This means the ring density peak may serve as a better metric for estimating the mass of a ring-forming planet from observations, where the Stokes numbers of the observed grains can be difficult to constrain. We apply a linear fit to our data and find that for all aspect ratios, the ring peak is located at approximately
\begin{equation}
    \label{eq:rpeak0}
    r_{peak} = (1.07 \pm 0.01)r_{p} + (5.3 \pm 0.3) R_{H}
\end{equation} 
which can be rearranged to solve for a planet's mass as
\begin{equation}
    \label{eq:rpeak}
    \frac{M_{p}}{M_{*}} = 3 \cdot \Bigg [ \frac{1}{5.3} \bigg(\frac{r_{peak}}{r_{p}} -1.07 \bigg) \Bigg]^{3}
\end{equation}

If it is assumed that the planet is located in the centre of an observed gap (of which the location is known) and the location of its associated ring is also known, this equation can be used to calculate the planet's mass. We note that the  slope of this linear fit broadly agrees with the measurements made by \citet{lodato2019}, and the ring inner edge and peak follow the same approximate scaling relation with  different offsets. The uncertainty in the gradient and offset in Equation \ref{eq:rpeak} primarily come from data points which deviate from the  linear trend due to the formation of ``shoulders'' or bumps just interior or exterior to the ring, which can be seen in Figure \ref{fig:1dsigmas}. The ring peak identification methods employed in this work always take the local density maximum to be the peak density but in a number of cases, it can be difficult to distinguish between what we define as a ``peak'' and what may be a ``shoulder'', leading to the anomalous data points seen in Figure \ref{fig:rpeaks}. \\

The origin of these "shoulders" is apparent by examining the full 2D images of our simulations, as shown in Figure \ref{fig:2dsigmas} in Appendix \ref{app1}. Radial asymmetries may be present in a ring but are not necessarily present at all azimuths. These effects may not appear in 1D azimuthally averaged dust density profile due to being cancelled out or smoothed over by the averaging of the azimuthal variations, but may still shift the measured location of the ring in the 1D profile used for ring fitting by widening the perceived ring peak. \\

The fact that the ring inner edge location being dependent on the Stokes number, while the ring peak is not, can be understood in terms of how the dust responds to the behavior of the gas. The ring peak location is set by the location of the pressure maximum in the gas, leading to it being the same for all Stokes numbers. The ring inner (and outer) edges are set by the interplay of the outward and inward flow of dust caused by the relative pressure gradients on either side of the pressure maximum in the gas. Different sized dust reacts to non-zero pressure gradients differently -- dust radial velocities for the same pressure gradient are higher for larger Stokes numbers (see Equation \ref{eq:vr}) -- which can lead to different inner and outer edge locations for different Stokes numbers. Larger dust grains also respond to changes in the gas on shorter timescales, although, as shown in Figure  \ref{fig:timeevo}, the dust rings in our models appear to have reached an approximately steady state by 1500 orbits. \\

\subsection{Ring masses}
\label{sec:results_rmass}

\begin{figure}
    \centering
    \includegraphics[width=1\linewidth]{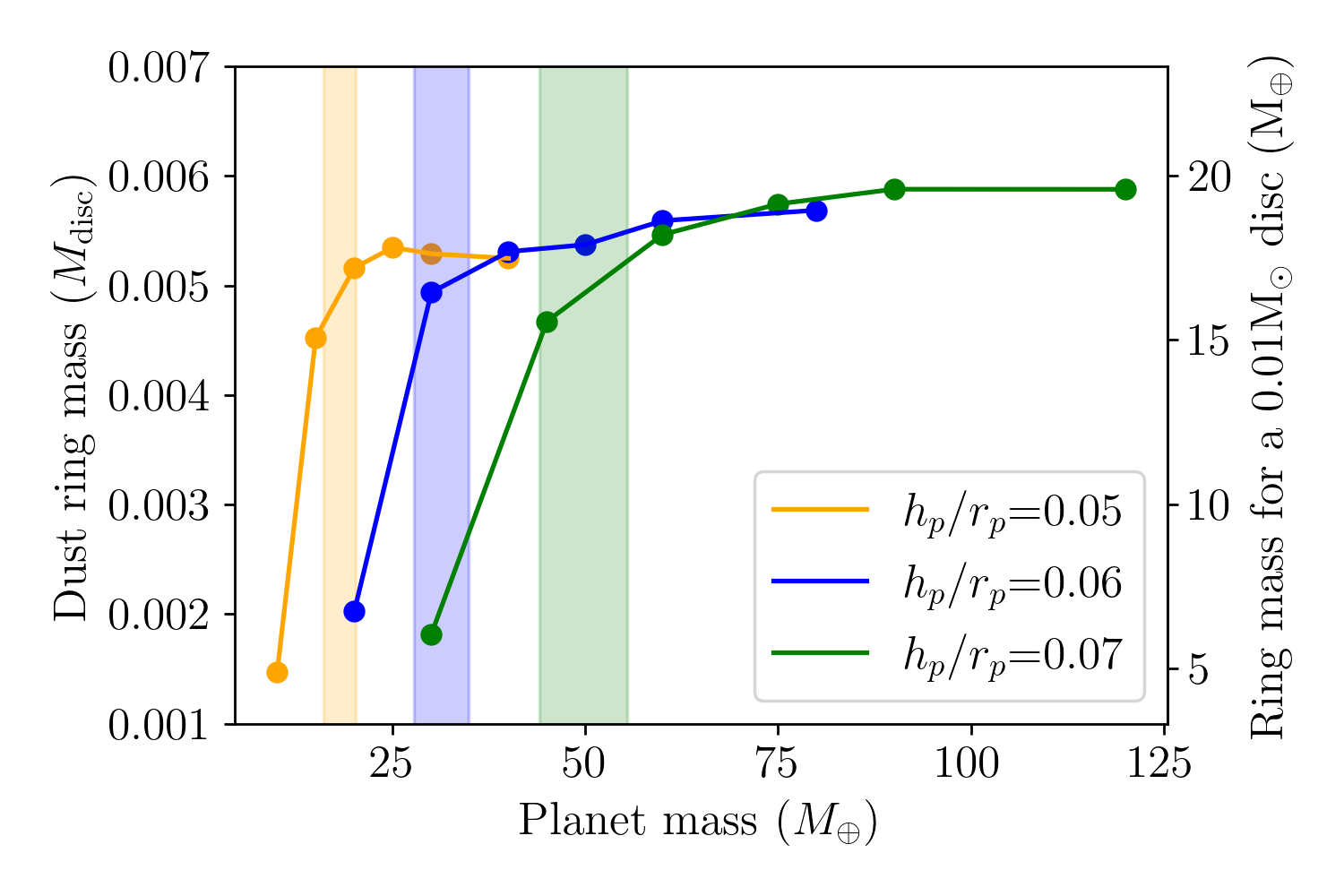}
    \caption{Mass contained in dust rings against planet mass for all three aspect ratios. The shaded regions indicate the pebble-isolation mass for each aspect ratio.}
    \label{fig:rmasses}
\end{figure}
Figure \ref{fig:rmasses} shows the mass contained within the rings for each planet mass. These masses are computed assuming the grains follow the MRN size distribution with a total dust-to-gas mass ratio of 0.01, as explained in Section \ref{sec:dmass_calc}. We find that the ring masses increase with planet mass for lower mass planets but plateau for planet masses greater than the pebble-isolation mass. This is consistent with our understanding of why ring widths also become fixed for planet masses greater than the pebble-isolation mass: lower mass planets produce leaky, weak dust traps which are seen as diffuse, wide rings with limited mass trapped in them. In contrast, planets with masses greater than the pebble-isolation mass trap dust more effectively, resulting in compact and dense rings, which are narrower than those produced by planets with masses below the pebble-isolation mass but also contain more material. \\

We find that if we scale our dimensionless ring masses such that we assume the total disc mass to be 0.01$M_\odot$ ($\simeq 330 M_\oplus$), the mass of solid material trapped in the dust rings is on the order of 20 Earth masses ($\sim 7$\% of the total dust mass in the disc) for the highest mass planets, consistent with the findings of \citet{dullemond2018}. The region interior to the planet in these cases contains around $25\%$ of the total dust mass of the disc, compared to around $44\%$ of the dust mass in the lowest planet mass cases, showing the effectiveness of higher mass planets at stopping the inflow of dust. Although they comprise a small fraction of the total dust mass, this result shows that dust rings may serve as sites for the formation of additional planetesimals with masses that are either comparable to or less than the mass of the initial planet. We note that there is negligible difference between the mass in the dust rings for discs with different aspect ratios, suggesting that our approximation that 7\% of the total dust mass can be trapped in the ring by planets above the pebble-isolation mass holds true for all discs (with an $\alpha$ of $10^{-4}$). \\

\subsection{Dust-to-Gas Ratios and Planetesimal Formation in Rings}

One of the means by which planetesimal formation could proceed is the streaming instability, which requires that the dust-to-gas ratio reach a sufficiently high value. \citet{Li_Youdin2021} have shown that this threshold is a function of the Stokes number and a review by \citet{Birnstiel2024} combines this result with the effect of turbulence on this threshold (See Figure 11 in that paper) to define a parameterised threshold for gravitational collapse via the streaming instability. From these results and at our value of $\alpha = 10^{-4}$, clumping of dust will occur in simulations at dust-to-gas ratios as low as 0.01 for Stokes numbers above 0.01. \\

\begin{figure}
\centering
\includegraphics[width=\columnwidth]{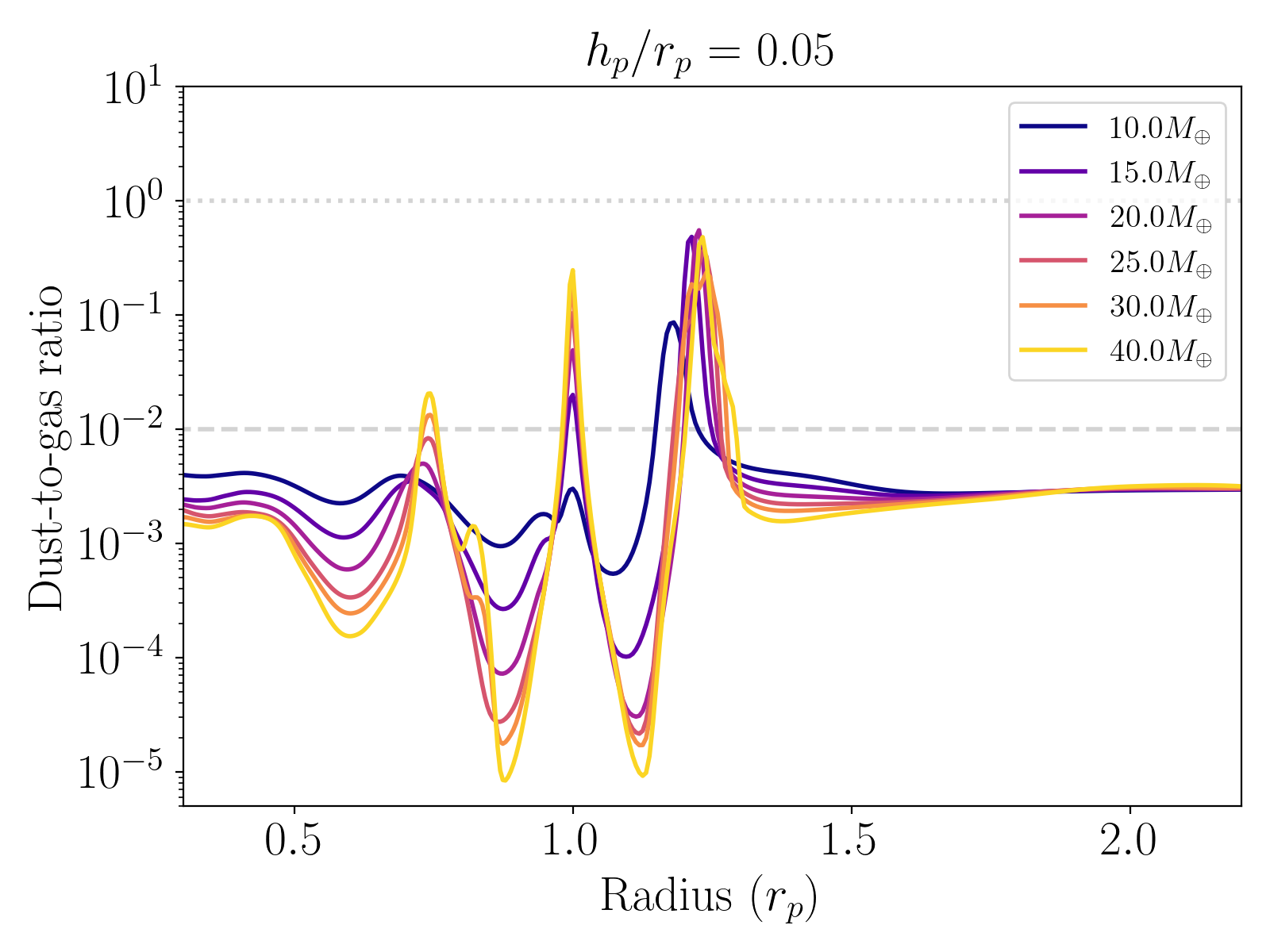}\par\vspace{1ex}
\includegraphics[width=\columnwidth]{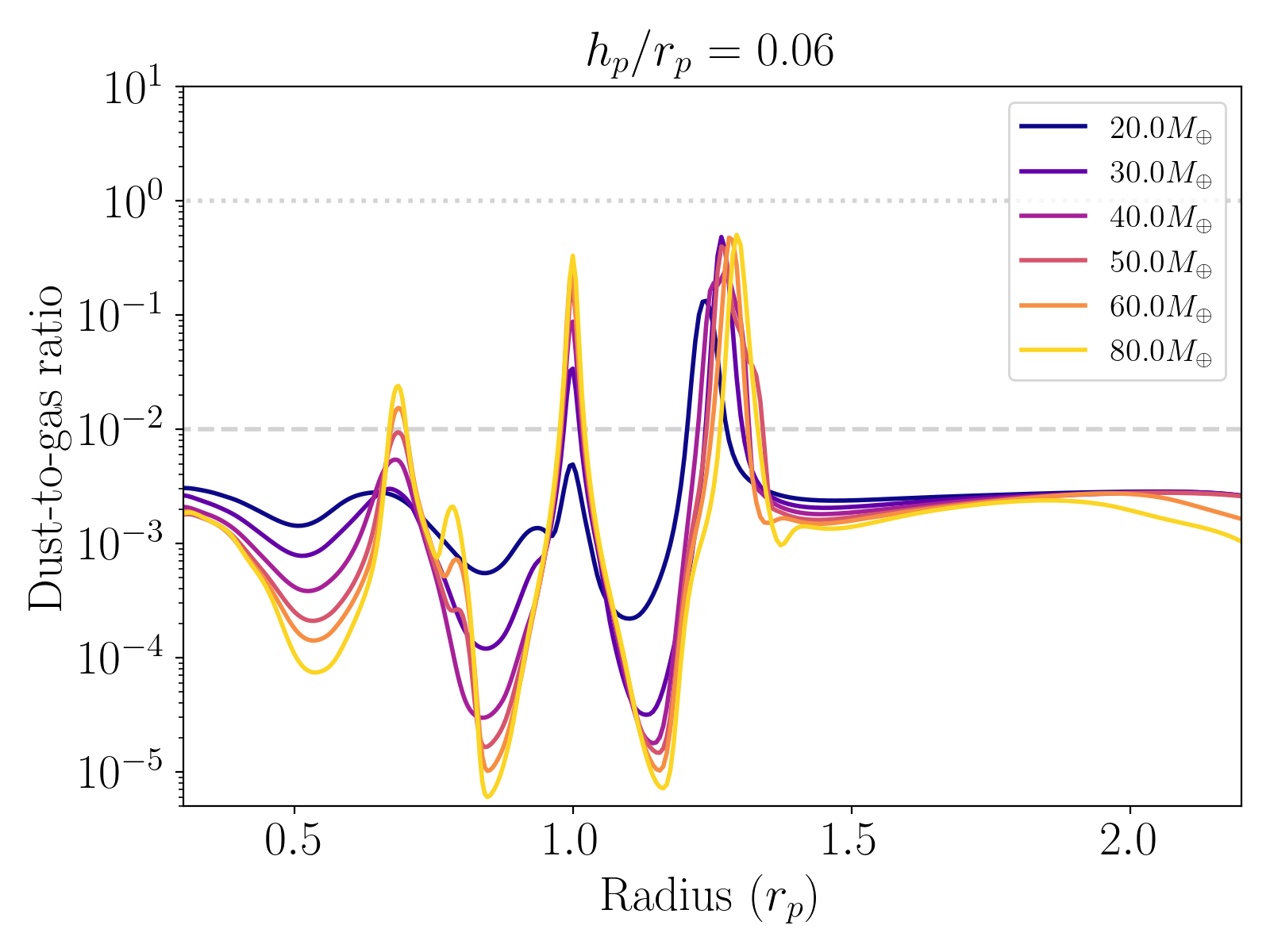}\par\vspace{1ex}
\includegraphics[width=\columnwidth]{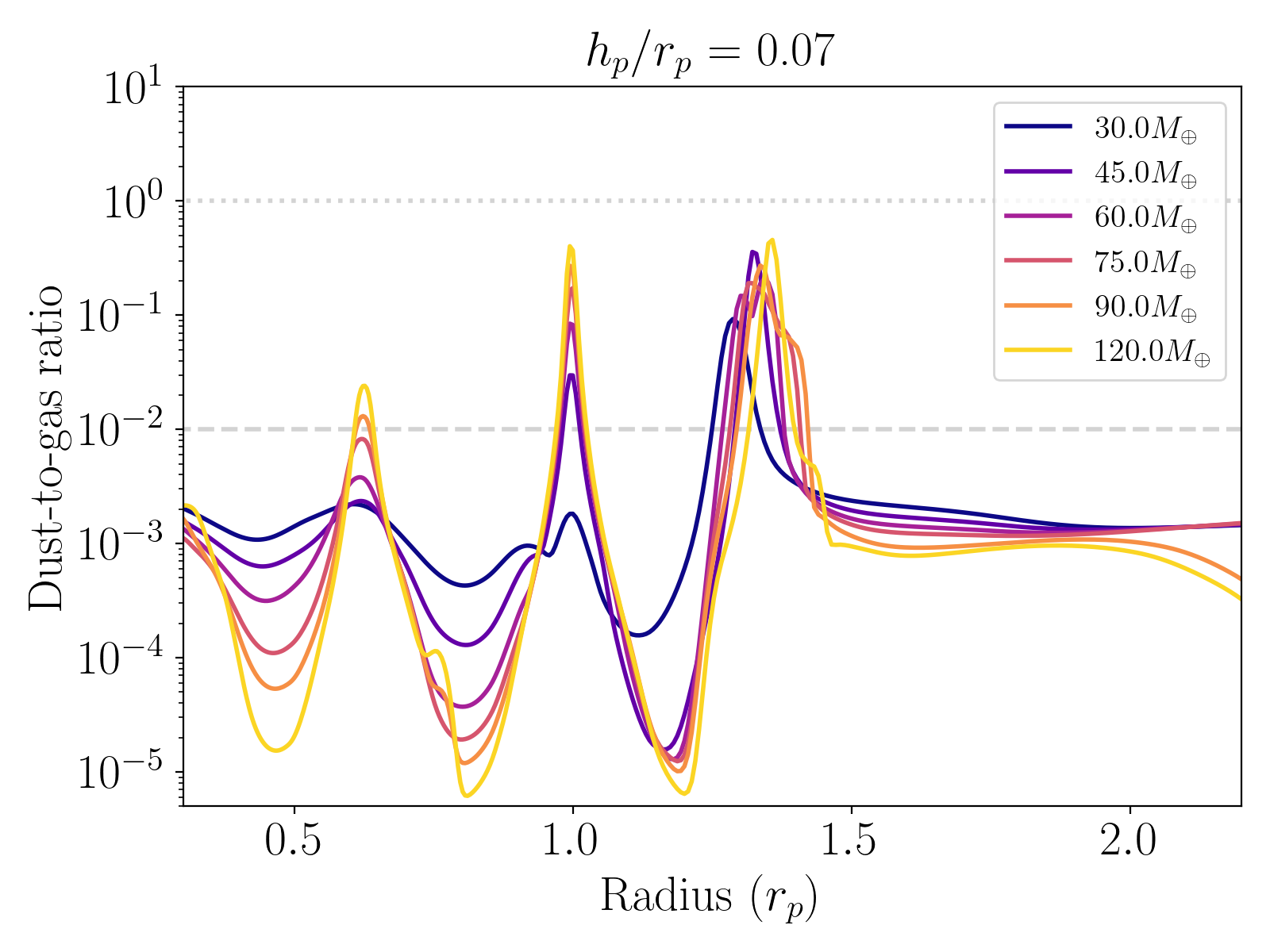}
\caption{Dust-to-gas ratios across the extent of the disc for all three aspect ratios at 1500 orbits. The dashed line indicates the initial dust-to-gas ratio of $10^{-2}$. The dotted line indicates a dust-to-gas ratio of 1, the value typically required to trigger the streaming instability.}
\label{fig:dgrs}
\end{figure}

In Figure \ref{fig:dgrs} we show the total dust-to-gas mass ratio across the radial extent of the disc for all three aspect ratios. We find that the dust-to-gas ratio is significantly enhanced in the dust ring, exceeding the criterion outlined by \citet{Birnstiel2024} for $\rm St \geq 0.01$ dust.  \\

\citet{lau2024} prescribed a probability distribution function for estimating the likelihood of gravitational collapse of dense filaments of material induced by the streaming instability. We compute the probability of such a collapse in our simulations using the following expression (equations 10 and 11 in \citet{lau2024}) :
\begin{equation}
\label{eq:ppf}
    \mathcal{P}_{pf} = \big[1+\exp\big(10(Q_{p}-0.75)\big) \big]^{-1}
\end{equation}

\begin{equation}
    Q_{p} = \sqrt{\frac{\delta}{\rm St_{avg}}} \frac{c_{s} \Omega}{\pi G \Sigma_{\rm d,local}}
\end{equation}\\

where $\delta$ is the small-scale diffusion parameter, for which we used $\delta=10^{-5}$, based on the results of \citet{schreiber2018}, $\rm St_{avg}$ is the weighted average Stokes number ($\rm St_{avg} = 0.023$), $\Omega$ is the angular velocity, $G$ is the gravitational constant, and $\Sigma_{\rm d,local}$ is the weighted average dust surface density in the ring. The weights applied for averaging the Stokes numbers and dust surface densities are the same as those described in Section \ref{sec:dmass_calc}.\\

\begin{figure}
    \centering
    \includegraphics[width=1\linewidth]{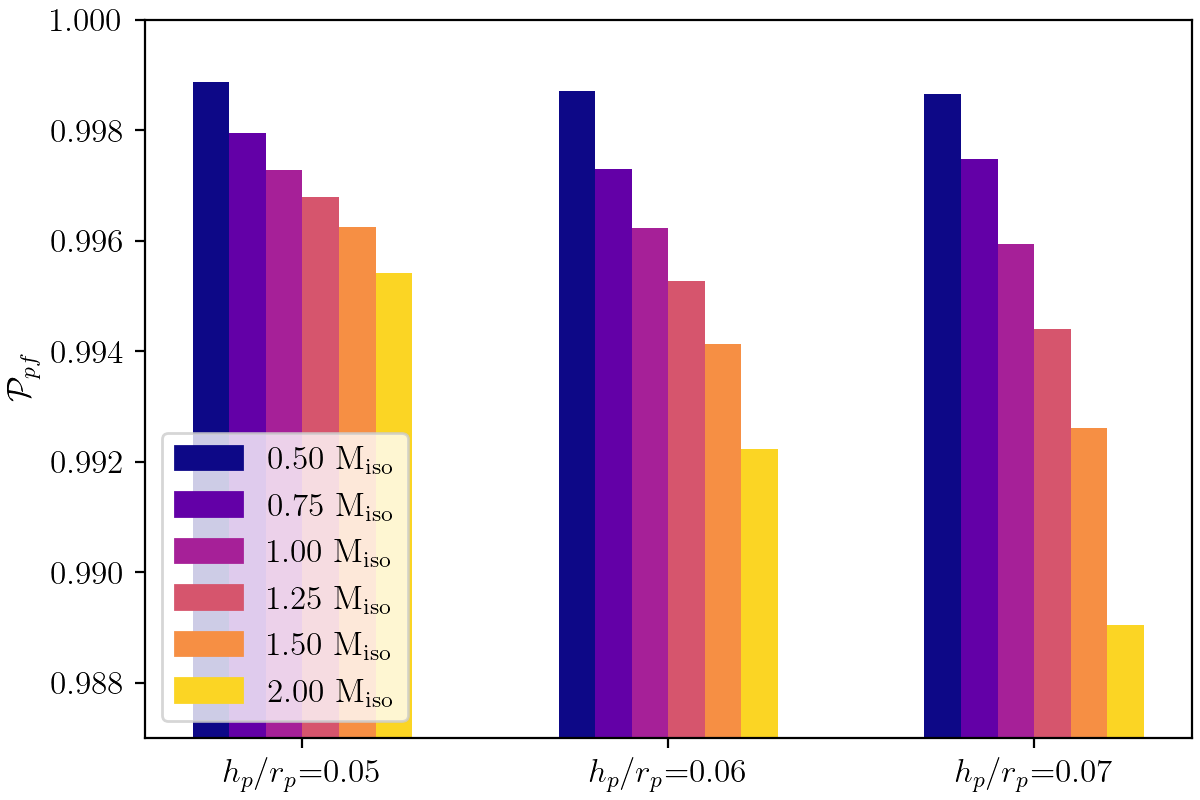}
    \caption{Probability of planet formation through the gravitational collapse of a clump formed via the streaming instability at the location of the ring peak, for all three aspect ratios, at 1500 orbits. All models showed a high likelihood of being capable of triggering the streaming instability in a ring.}
    \label{fig:ppfs}
\end{figure}

In Figure \ref{fig:ppfs} we plot the probability of planet formation that results from this calculation.  Surprisingly, all scenarios predict a high probability of inducing gravitational collapse (at least 98\% for all aspect ratios). This result is likely enhanced by the absence of dust growth and fragmentation in our models - processes which would limit the amount of dust trapped in a ring by enabling trapped dust grains to fragment to smaller sizes which could then escape the dust trap (Faruqi et al. in prep). As a result, we conclude that Equation \ref{eq:ppf} is more applicable in models that do not include growth. \\

We also note that across all aspect ratios, this probability drops off with increasing planet mass, even for planets heavier than the pebble-isolation mass, in spite of the mass and width of rings produced by planets heavier than the pebble-isolation mass being constant. This is due to the radial location of the ring changing with planet mass. As the planet mass increases, the ring is pushed radially outwards, meaning the disc scale height at the ring's location increases (for a disc with a flaring index of 0.25), spreading dust in the vertical direction, while the local dust surface density also decreases. The compounding effect of these two factors decreases the likelihood of gravitational collapse for higher mass planets.\\

We conclude that dust traps formed by planets have the potential to be highly probable sites for the formation of future planetesimals, which raises the question of why we do not see more evidence of planets forming in rings. Factors inhibiting the growth of dust in rings may offer an explanation on this, and will be the subject of study of a future publication (Faruqi et al. in prep). \\


\subsection{Refining our understanding of pebble-isolation theory}
\label{sec:understanding_miso}


\begin{figure}
\centering
\includegraphics[width=\columnwidth]{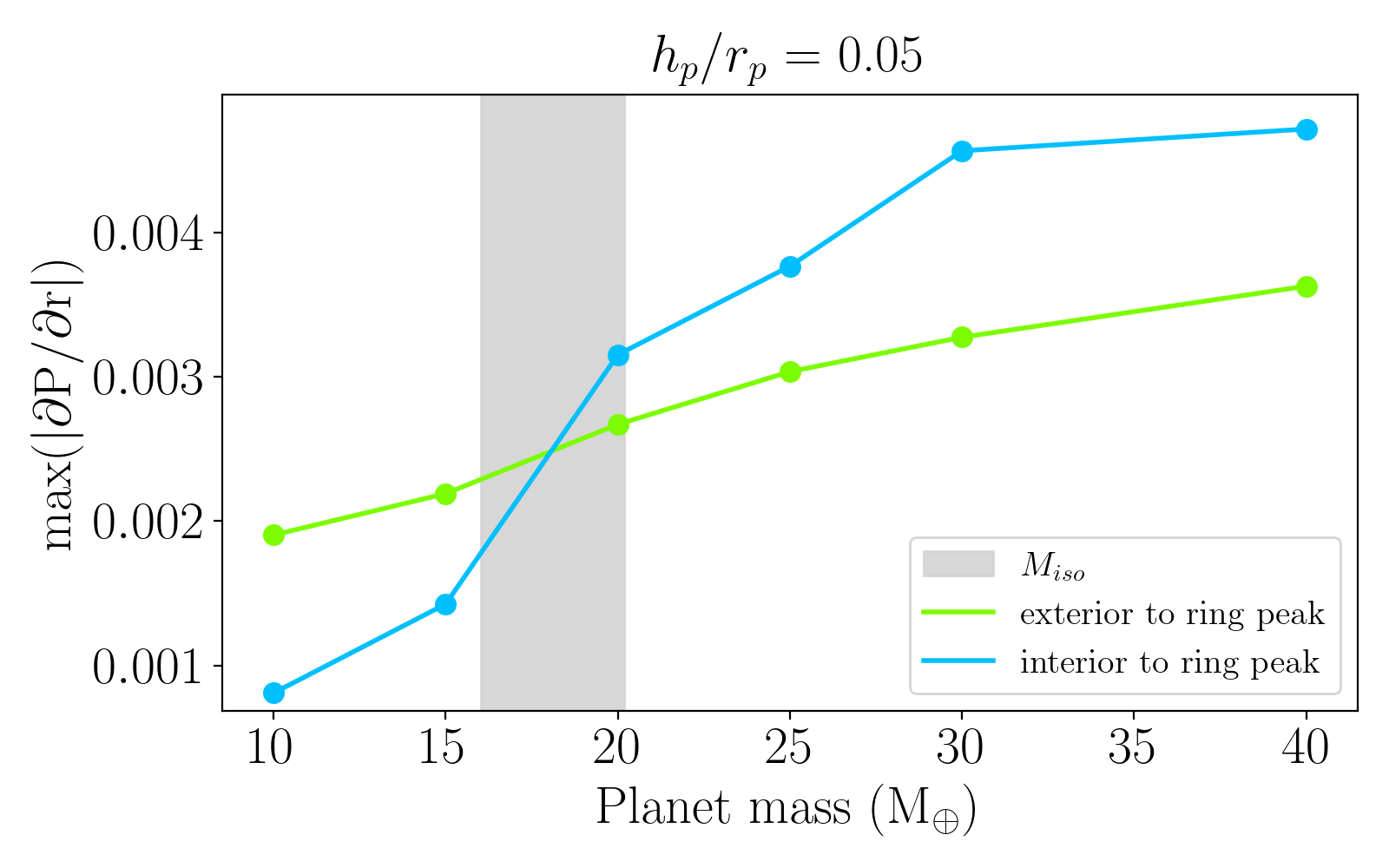}\par\vspace{1ex}
\includegraphics[width=\columnwidth]{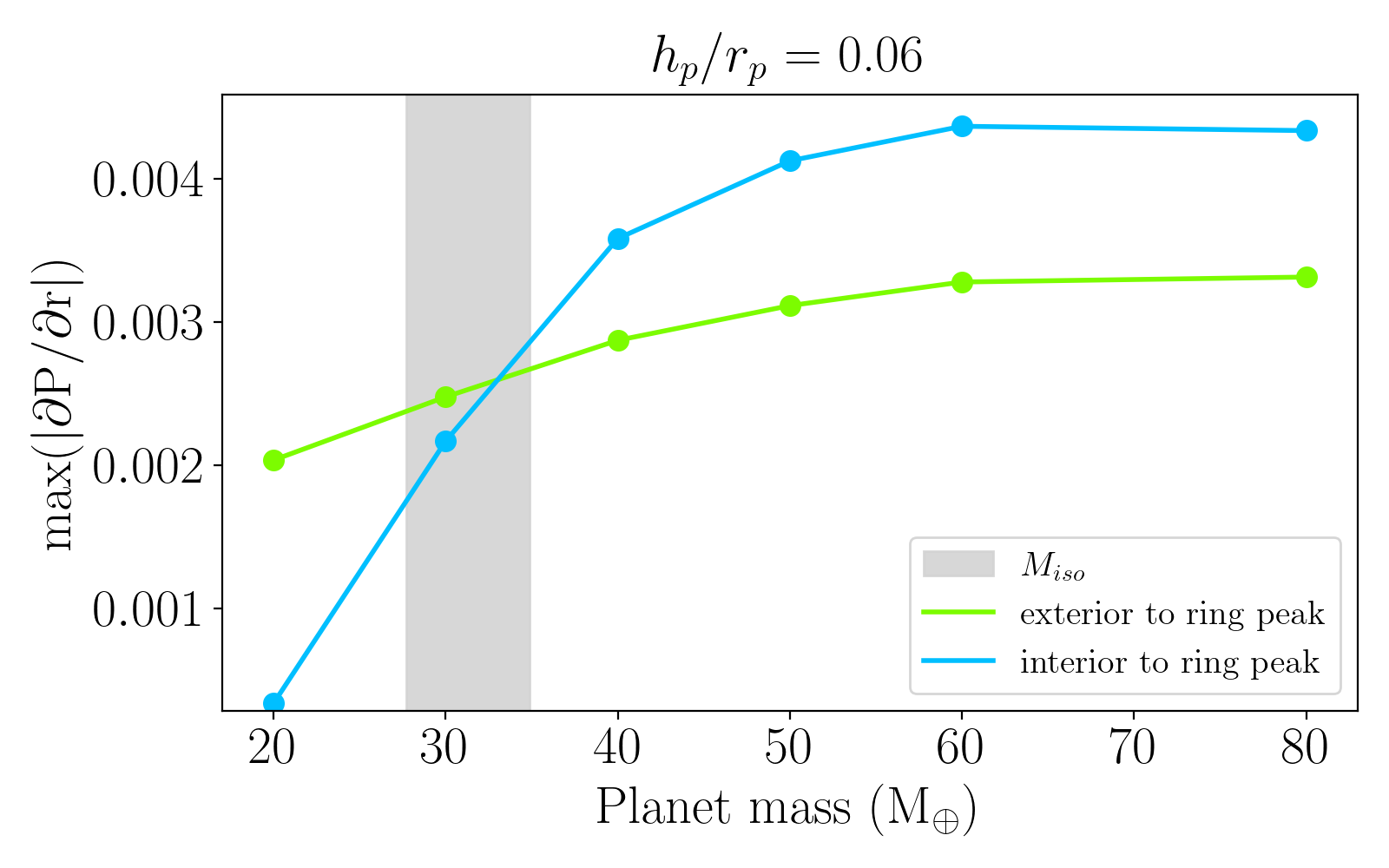}\par\vspace{1ex}
\includegraphics[width=\columnwidth]{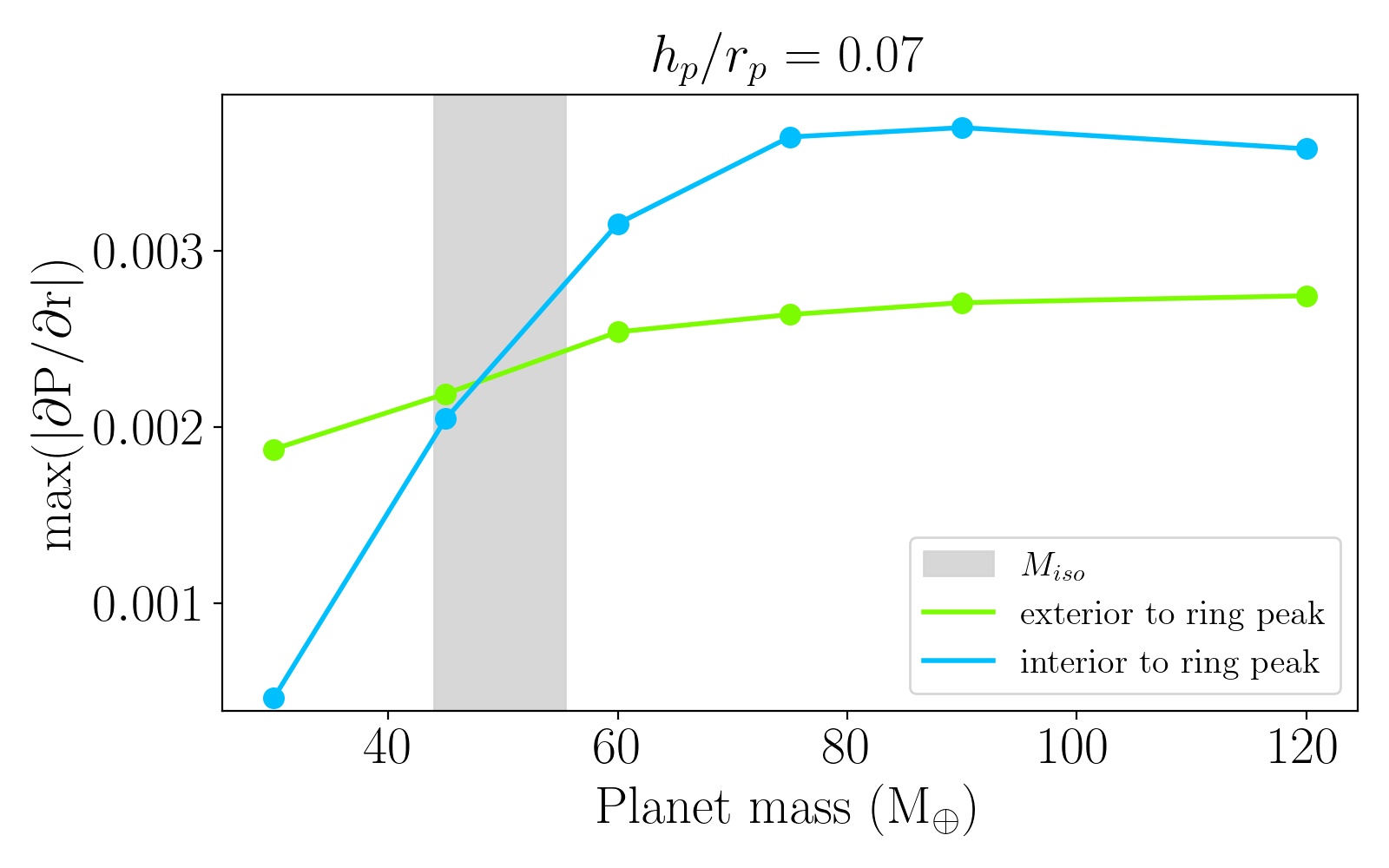}
\caption{Absolute values of the maximum pressure gradient interior and exterior to the pressure maximum plotted against planet mass for all three aspect ratios. The gray box indicates the range of values for the pebble-isolation mass for the Stokes numbers modeled.}
\label{fig:dpdrs}
\end{figure}

The difference in the behavior of dust when the perturbing planet's mass is above or below the pebble-isolation mass can be interpreted qualitatively in terms of the mechanisms that lead to the formation of a dust trap. Dust grains in the outer disc undergo radial drift, following the negative gas pressure gradient until they reach the pressure maximum formed by a perturbing planet.  On the other side of the pressure maximum is a positive pressure gradient, expelling material radially outwards in the disc. Given that the radial velocity of grains is proportional to $\partial P/ \partial r$ (see Equation \ref{eq:vr}), the flow of material in either direction competes due to the opposing pressure gradients on either side of the pressure maximum. The direction of the net flow of dust is therefore determined by the relative magnitudes of the pressure gradients, which then determines the region of space in which the dust is able to be trapped. \\

In Figure \ref{fig:dpdrs}, we plot the magnitudes of the maximum pressure gradients both interior and exterior to the pressure maximum. We find that the exterior pressure gradient (pushing material inwards) exceeds the interior pressure gradient for planet masses less than the pebble-isolation mass. This means that lower mass planets produce pressure bumps that are insufficient in their ability to compete with the radial drift of dust grains and thus, allow material to leak out of the weak dust trap via the ring inner edge, resulting in a wider ring. However, planets with masses greater than the pebble-isolation mass are able to expel more material than is drifting inwards, by producing a steeper positive pressure gradient interior to the pressure maximum. Unlike the first scenario, this does not lead to a widening of the ring in the exterior direction, since material that is pushed outwards rapidly enters a region with a negative pressure gradient, causing its direction to be reversed once again. The radial location in the disc at which this occurs (i.e. the ring outer edge) depends on the relative values of the interior and exterior pressure gradients.\\

Figure \ref{fig:dpdrs} also shows that the ratio of the interior to exterior pressure gradients becomes approximately constant with increasing planet mass, for planet masses greater than the pebble-isolation mass. This explains why the ring width also stops changing past this point, since the location of the ring outer edge relative to the ring peak becomes fixed and hence, the distance between the outer and inner edges becomes approximately constant.  \\

In past literature, the pebble-isolation mass has been quantified empirically, based on detailed modeling across a vast parameter space. This has produced varying prescriptions for calculating the pebble-isolation mass \citep[e.g.][]{lambrechts2014, ataiee2018, bitsch2018}, all of which parameterize it in slightly different ways. Here, we propose another definition for the pebble-isolation mass: it is the minimum planet mass which perturbs the gas enough for the pressure gradient interior to the pressure maximum to exceed the pressure gradient exterior to it.\\

This definition is in agreement with Equation \ref{eq:miso}, derived by \citet{bitsch2018}, for the parameter space explored in this work. However, the expression given by \citet{bitsch2018} is better suited for theorists as a means of selecting input parameters for a simulation in which the goal is to model ring formation. It is not as helpful for estimating planet masses based on observed ring properties. \\

An important advantage of our definition is that it can be more directly linked to observations. If the gas surface density profile and temperature profile of a disc with a known gap can be approximated from observations of its gas component, a comparison of the gas pressure gradient on either side of the observed pressure maximum could allow us to constrain whether or not the planet creating the pressure bump falls above or below the pebble-isolation mass for that disc. The exact values of the pressure gradient would not need to be known, they would only need to be measured with a small enough margin of error to be able to say which is greater in magnitude. Based on this, the mass of a planet embedded in a disc could be constrained without the dust component of the disc even being observed directly.  \\

\section{Discussion}
\label{sec:discussion}
\subsection{Observational implications}

So far we have outlined the following ways in which observable characteristics of rings can be related to the mass of a planet:
\begin{enumerate}
    \item Measured ring widths could yield estimates of the pebble-isolation mass and possibly the planet mass.
    \item The ring peak location can be used to estimate planet mass, assuming the planet's location is known. This does not require knowledge of the disc's aspect ratio and works for all observing wavelengths. 
    \item The pressure gradients interior and exterior to a pressure maximum can be used to determine whether the planet mass is above or below the system's pebble-isolation mass. 
\end{enumerate}

However, the application of these results requires some prerequisite knowledge of the disc's geometry and properties such as the viscosity. For instance, the relationship we find between planet masses and ring widths holds for all aspect ratios but the absolute values that can be derived from a measurement of the ring width differ depending on the aspect ratio. We also only tested a single value of the $\alpha$-viscosity but it is known that substructures such as gaps and rings are harder to form in more viscous discs. In order to estimate planet masses from ring widths, observers would need therefore need to have a prior estimate of the disc's aspect ratio and viscosity, which is rapidly becoming easier to do \citep{exoalmaV2025, exoalmaXVI2025}. If the aspect ratio and the  $\alpha$-viscosity of a disc can be measured, then the pebble-isolation mass for a protoplanetary disc can be calculated using the prescription outlined in \citet{bitsch2018}. Models generated for the determined aspect ratio and the approximate Stokes number that corresponds to the wavelength of the observations (which can be determined if the gas surface density profile is known) can then yield the equivalent of Figure \ref{fig:rwidths} for an observed system. For observers, it would be ideal to conduct observations at smaller wavelengths, as the variation of ring widths with planet masses are most prominent for the smallest Stokes values.\\

If the planet's mass in the system falls below the pebble-isolation mass, the observed ring width could then be used to estimate the planet's mass. This estimate can then be consolidated using other methods for estimating the masses of planets embedded in discs, such as those that use the gap depth or width, as defined by \citet{kanagawa2016}, or by using the ring peak location via the methods we describe in Section \ref{sec:results_rpeak}.  On the other hand, if the ring width measured is approximately equal to the value predicted for planet masses greater than the pebble-isolation mass, this would not allow us to estimate the planet's mass.  However, it can still provide a lower bound on the planet's mass.\\

Observations at any wavelength within the sub-mm to mm range studied here could be used to resolve the location of the brightest annulus in the dust ring, which is independent of Stokes number. This could be used to estimate the mass of the putative planet causing it, under the assumption that the planet's orbital radius is the center of the observed gap. The ring peak location is an especially useful metric for estimating planet masses since we see that, unlike ring widths, it does not plateau for planet masses above the pebble-isolation mass. \\

These methods can be used in conjunction with other methods for estimating planet mass, such as those discussed in \citet{kanagawa2016}, which use gap widths and depths to estimate planet masses, or the detection of velocity kinks \citep{pinte2018, teague2018}. If two or more methods yield approximately the same planet mass, it can be safely concluded that (a) the gaps and rings seen are indeed caused by a planet, confirming the presence of a planet in a disc, and (b) we know the mass of the planet, which can then be used to inform our understanding of planet formation processes in protoplanetary discs and how the architecture of planetary systems relates to the characteristics of the discs they result from. Conversely, if the methods yield significantly different planet masses, this may indicate that the pileup are the result of a different mechanism, such as MHD winds or material crossing a snow line. If that is the case, it would still provide useful insight into the conditions planets may be forming in.\\ 

\subsection{Application of results to the PDS 70 system}\label{subsec:pds70}
\label{sec:pds70}
\begin{figure}
    \centering
    \includegraphics[width=1\linewidth]{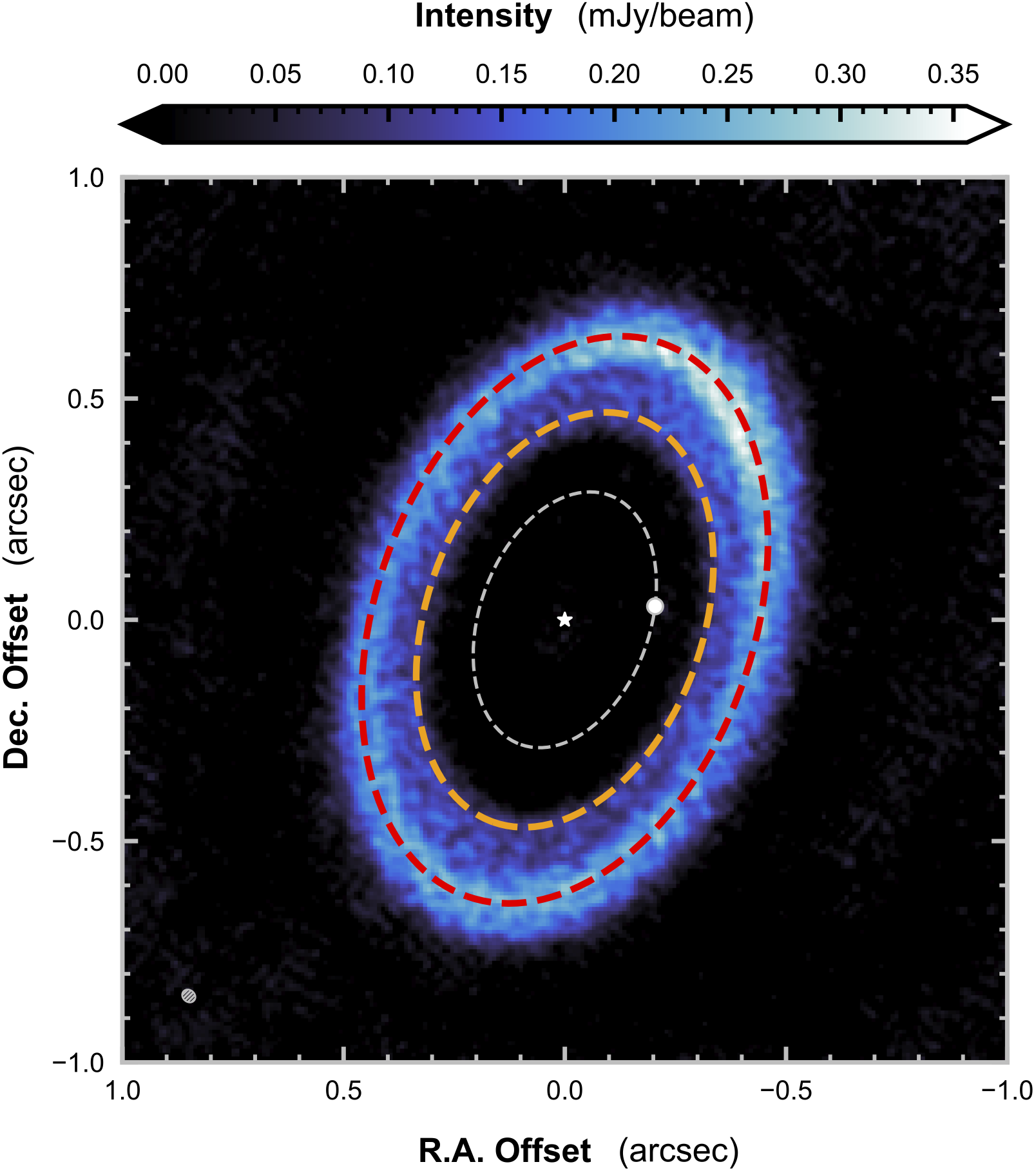}
    \caption{For illustrative purposes, a map of the dust continuum emission in the PDS 70 disk \cite[ALMA pipeline image in Band 7 from program 2018.A.00030.S;][]{benisty2021}. The dashed orange and red ellipses indicate the radial locations of the ring's shoulder and outer peak that we adopt for our discussion in Section \ref{subsec:pds70} (54.5 AU and 74.5 AU in deprojected radius, respectively). The white dashed ellipse represents a circular orbit for PDS 70 c with a deprojected radius of 33.6 AU. All three circular trajectories are projected onto the sky using an inclination of $49.9^{\circ}$ and position angle of $160.5^{\circ}$ \citep{Fasano2025}, adopting a distance of $112.4$~pc \citep{gaia2021}.}
    \label{fig:pds70}
\end{figure}
To demonstrate the validity of our results, we apply them to observational data to estimate the mass of one of the two confirmed planets in the PDS 70 system and, in the following section, constrain the planet masses that could be responsible for the formation of rings in the exoALMA sample. \\

The PDS 70 system consists of a single T Tauri star with a protoplanetary disc that has two confirmed planets, PDS 70b and PDS 70c. \citet{Hammond2025} recently used a range of differential imaging techniques to calculate the mass of PDS 70, as well as the orbital radius and mass of the two confirmed planets hosted by the system. For the outermost of the two planets, which would be the primary perturbing mass shaping the outer ring, they calculate its orbital radius to be $33.6^{+2.4}_{-2.6}$ AU. They calculate the mass of PDS 70 to be $0.92 \pm 0.07 M_\odot$. ALMA Band 6 and 7 observations of the system have shown a bright outer ring of dust, located at $\sim 72-77$ AU, with a narrower ring interior to it, located at $\sim 53-56$ AU \citep{Fasano2025}. By combining the location of the ring,  measured by \citet{Fasano2025}, with the estimates provided by \citet{Hammond2025} for the mass of PDS 70 and the orbital radius of PDS 70c, we can use Equation \ref{eq:rpeak} to estimate the mass of PDS 70c, assuming that PDS 70c is solely responsible for the formation of the ring.\\

The multi-peaked nature of the ring in PDS 70 makes it difficult to determine which ring location to use for this calculation. As described in Section \ref{sec:results_rpeak}, Equation \ref{eq:rpeak} already accounts for the possibility of non-Gaussian ring profiles, which may contain multiples peaks or a single peak with additional ``shoulders''. Therefore, we apply our results to both the ring peak and the inner shoulder of the ring in PDS 70.\\ 

In Figure \ref{fig:pds70}, we illustrate the midpoint locations of the two ring peaks, as well as the orbital trajectory of PDS 70c (dashed white line), overlaid on the ALMA pipeline image of the disk in Band 7 from program 2018.A.00030.S \citep{benisty2021}. 
We first consider the location of the inner ``shoulder'' of PDS 70's ring at 54.5 AU \citep{Fasano2025} (dashed orange line) as being induced by the presence of PDS 70c, since it is closer to the planet's orbit. Based on this, we calculate the mass of PDS 70c to be $3.3^{+7.1}_{-2.4} M_{\rm Jup}$.  This aligns very well with the masses estimated by \citet{Hammond2025} ($5.4 \pm 3.1 M_{\rm Jup}$) and \citet{Ruzza2025} ($3.83^{+0.90}_{-0.66} M_{\rm Jup}$), both of which employed completely independent methods to those described here. \\

If we now repeat this calculation using the location of the 74.5 AU ring peak - illustrated  by the dashed red line \citep[the midpoint of the range provided by][]{Fasano2025} - we estimate the mass of PDS 70c to be $29.3^{+42.5}_{-17.5} M_{\rm Jup}$. This value is significantly higher than previous estimates, which range from as little as $1 M_{\rm Jup}$ \citep{Wang2020} up to $8.5 M_{\rm Jup}$ \citep[the upper limit of][]{Hammond2025}. This discrepancy could be due to the assumption that PDS 70c is the only mass perturbing the disc to form a ring, given that it is known that at least one other planet of comparable mass inhabits the system. It may also be related to the fact that Equation \ref{eq:rpeak} is based on models in which the planets are on perfectly circular orbits, whereas the planets in PDS 70 are believed to have non-zero eccentricities, which may lead to the formation of a ring further out than would be expected from a planet on a circular orbit. It may also be the case that the presence of the inner shoulder pushes the ring peak outwards slightly, as seen in our own results, making the inner shoulder's location more suitable for determining the planet mass.\\

\subsection{Application of results to exoALMA sample}

\begin{table*}[]
\centering
\begin{tabular}{|c||c|c|c|c|c|c|c|}
\hline
\multicolumn{1}{|c||}{System} & \multicolumn{1}{p{2cm}|}{Pressure max location (AU)} & \raisebox{-0.6\height}{\shortstack{Gas gap \\ location (AU)}} & \raisebox{-0.6\height}{\shortstack{Dust ring \\ location (AU)}} & \multicolumn{1}{c|}{Gas Tracer} & \multicolumn{1}{c|}{$M_{p} > M_{\rm iso}$} & Planet mass ($M_\oplus$)  & $h_{p}/r_{p}$  \\ \hline \hline
AA Tau  & 74.1-77.4  & $64.3 \pm 5.1$  & $71.7 \pm 6.1$  & 12CO & Y & $0.5^{+21.2}_{-0.5}$      & $\lesssim 0.05$ \\ \hline
DM Tau  & 85.4-90.7  & $71.8 \pm 4.1$  & $89.5 \pm 6.5$  & 13CO & N & $16.6^{+59.0}_{-15.5}$    & $\gtrsim 0.02$ \\ \hline
J1615   & 116.8-135.4 & $82.8 \pm 8.8$ & $105.6 \pm 8.8$ & 12CO & N & $69.3^{+240.6}_{-64.3}$   & $\gtrsim 0.03$ \\ \hline
LkCa15  & 99.1-105.1 & $86.3 \pm 5.9$ & $99.5 \pm 5.1$ & 12CO & N & $4.5^{+28.1}_{-4.4}$      & $\gtrsim 0.01$ \\ \hline
V4046 Sgr & 28.2-31.2 & $20.4 \pm 2.5$ & $27.2 \pm 3.2$ & 12CO & Y & $215.7^{+894.0}_{-206.8}$ & $\lesssim 0.2$  \\ \hline
\end{tabular}
\caption{Inferences made using the methods described in Section \ref{sec:results} for 5 discs in the exoALMA sample. Column 6 uses the disc's gas pressure profile to constrain the masses of the presumed planets in each system relative to the system's pebble-isolation mass. Column 7 contains the masses estimated using Equation \ref{eq:rpeak} based on the dust gap and ring locations in \citet{Stadler2025}, which are listed in columns 3-4. Column 8 combines these two sets of inferences to constrain each disc's aspect ratio.}
\label{tab:exoalma}
\end{table*}

We also  apply our result relating the  planet mass to the magnitude of the gas pressure  gradients on either side of a pressure maximum (see Section \ref{sec:understanding_miso}) to the exoALMA sample of discs. In \citet{Stadler2025}, the  authors present gas pressure gradients for 9 out of the 15 discs in the sample (6 discs in the sample were omitted due to a lack of the necessary data for calculating pressure gradients). We use the pressure gradients presented in Figures 7, E7, and E8 in \citet{Stadler2025} to estimate whether or not the the planet that may be inducing a pressure bump has mass that falls above or below the pebble-isolation mass for that disc, according to our findings described in Section \ref{sec:understanding_miso}. We also use the dust gap and ring locations in \citet{Stadler2025} to estimate the planet mass, using Equation \ref{eq:rpeak}.  \\

For each disc, we identify a pressure maximum wherever the sign of the pressure gradient changes from positive to negative with increasing radius. For each of the pressure maxima identified, we then compute the maximum value of $\partial P/ \partial r$ within 5 AU of the pressure maximum in either direction and compare these values to infer the upper or lower bound on the planet mass that may have caused it. Given that the pebble-isolation mass spans a range of planet masses for differing Stokes numbers, we require that that difference between the pressure gradients on either side of a pressure maximum must be greater than or equal to 0.001 (the minimum difference between the interior and exterior pressure gradients for planet masses above or below the pebble-isolation mass in Figure \ref{fig:dpdrs}) to make an inference about the planet's mass. We check the  data provided for both 12CO and 13CO. To account for uncertainties in the stellar mass estimates used, we consider the range spanned by a $\pm 3 \%$ difference in the stellar mass, which would shift the entire pressure profile up or down, when identifying co-located pressure maxima and rings. Based on this variation, we provide ranges, rather than exact values, for the locations of the pressure maxima found. \\

We then identify dust gap and ring pairs such that the ring location falls within 1.5 times the gap location, since this is where we expect the ring to be, based on \citet{rosotti2016minimum} and our own results. We expect dust rings and the gas pressure maxima that create them to reside at approximately the same location in the disc. Therefore, we limit our selection of pressure maxima to those with a ring identified at its approximate location: this leaves us with 5 systems, with one pressure maximum each. We use the ring and gap locations, along with the stellar masses in \citet{izquierdo2025}, to estimate the masses of the presumed planets forming the rings. We then combine these planet mass estimates with our inferences on how the planet mass compares to the system's pebble-isolation mass (based on the pressure gradients, according to the methods described in Section \ref{sec:understanding_miso}) to apply an upper or lower bound on the disc's aspect ratio. The pressure gradients around the pressure maximum tell us whether the planet mass we have estimated using Equation \ref{eq:rpeak} is greater than or less than the pebble-isolation mass. If we assume the disc's $\alpha$-viscosity to be $10^{-4}$, the disc's surface mass density slope and flaring index to be 1 and 0.25 respectively, and assume a fixed Stokes number (here, we use 0.01), Equation \ref{eq:miso} can be combined with Equation \ref{eq:rpeak} and solved to obtain an inequality that constrains the aspect ratio of a protoplanetary disc i.e.
\begin{equation}
    3 M_{*} \Bigg [ \frac{1}{5.3} \bigg(\frac{r_{peak}}{r_{p}} -1.07 \bigg) \Bigg]^{3} \lessgtr \bigg(25 + \frac{\alpha/2 \rm St}{0.00476} \bigg)f_{\rm fit}  M_\oplus\\
\end{equation}

where $f_{\rm fit}$ is as defined in equation \ref{eq:ffit} and is proportional to $h_{p}/r_{p}$. The choice of inequality sign will depend on whether the pressure gradients indicate that the planet mass falls above or below the system's pebble-isolation mass. This inequality can then be solved to apply an upper or lower limit on the aspect ratio. We acknowledge that observed discs may have slopes or flaring indexes that differ from the values we have assumed in our calculations, which would slightly change the value of $f_{\rm fit}$ on the right-hand side of this inequality, but should have a negligible effect compared to the uncertainties on the pressure gradients. Our findings from these calculations are summarized in Table \ref{tab:exoalma}.\\

We identify 5 potential planets within the exoALMA sample, 2 of which are expected to fall above the pebble-isolation mass of the system. Based on the planet masses estimated for these systems, we apply upper bounds on the aspect ratios of these systems. It should be noted that the uncertainties on the planet mass estimates are large due to the exponent of 3 in Equation \ref{eq:rpeak}, which means that small uncertainties in the ring locations translate to large uncertainties in the planet mass. As a result, the bounds we place on the aspect ratios are also quite broad.\\

We find that AA Tau likely has an aspect ratio that is lower than the canonical value of 0.05, since the planet mass we estimate for it is low but is predicted to be above the pebble-isolation mass for the system. This result comes from the fact that the identified dust ring sits unusually close to its associated gap, where we assume the planet is located. It is possible that the planet actually sits closer to the inner edge of the gap, which would increase the planet mass estimate to a value more in line with expectations.\\

Conversely, V4046 Sgr suggests a much higher mass for a planet that is also above the pebble-isolation mass, leading to an aspect ratio that may be higher than what we would typically expect. This is explained by the system's stellar mass being high ($1.76 M_\odot$). The remaining 3 systems, all of which have presumed planets with masses below the pebble-isolation mass, have lower limits on their aspect ratios that are in agreement with typical expectations of disc aspect ratios \citep[e.g.][]{armitage2019}. \\

Of the systems we apply our results to here, only one that we know of has had its planet mass estimated in previous work. \citet{Ruzza2025} estimated the mass of the planet in DM Tau to be $12.7^{+3.2}_{-6.3} M_\oplus$, which aligns closely with our result for that system. There are currently no other estimates of the planet masses or disc aspect ratios for the remaining 4 systems discussed here so the accuracy of these results remains to be verified. The purpose of this section is largely to demonstrate how these results may be applied, particularly as more observational data of disc pressure profiles and ring and gap structures become readily available through future data releases. \\

\subsection{Caveats}
Here, we discuss the mechanisms omitted from our models and how they may have influenced our results. The primary simplification in our models is that all planets are on fixed orbits, unable to migrate. For the least massive planets modeled here, including migration would most likely not have a significant effect on the results, as the type I migration rate for low mass planets is relatively slow compared to higher mass planets. However, for the more massive planets, the inwards migration would change the magnitude of the torque exerted by the planet on the material exterior to its orbit, which would alter the morphology of the rings. \citet{meru2019ring} studied the impact of planetary migration on rings and found that a migrating planet can produce a gap and pressure maximum with a trailing outer edge, resulting in the strength of the dust pileup and depth of the gap being reduced by a factor of a 2-3. \\

We do not include dust feedback onto the gas component of the disc, which becomes a significant effect once the dust-to-gas ratio reaches unity. As seen in Section \ref{sec:results_rmass}, although our models do not produce dust-to-gas ratios quite as high, they are over an order of magnitude higher than the initial value of 0.01, meaning that the behavior of the gas may have been altered by dust feedback, if it had been included. Based on the work of \citet{kanagawa2018}, we expect that the main consequence of this would have been to reduce the gas pressure gradient, allowing more dust to escape the dust trap and potentially widening all rings. In a similar vein, \citet{huang2025} demonstrated that modeling dust traps in three spatial dimensions rather than two produces leakier dust traps, which is not captured by our 2D models. \\

Our models do not include collisional growth of dust grains; we assume fixed populations of dust across 5 discrete Stokes values. In reality, dust trapping does influence the rate of dust growth, particularly in the rings, but does not cause the overall size distribution of grains to deviate strongly from the MRN distribution, hence including growth in these models is unlikely to change our results significantly. \\

We did not consider the possibility of multiple ring structures formed by a single planet. It has been shown that planets embedded in low-viscosity discs \citep[e.g.][]{Zhu2014,Bae2016,Dong2018} can produce multiple concentric rings, with the additional ring(s) being located at much higher radii than the planet's orbital radius. However, it is likely that even if the methods outlined in this paper were applied to secondary rings, they would generate inconsistent estimates for the planet mass, allowing observers to rule out an embedded planet being the origin of such rings. Since we did not run models at low enough viscosities to test this, we cannot confirm whether or not our results would hold for the innermost ring in a series of concentric rings i.e. the first one exterior to the planet's orbit. \\

\section{Conclusion} \label{sec:conclusion}
In this paper, we present the results of a series of 2D hydrodynamical simulations of dust and gas in a protoplanetary disc with an embedded planet. We explore how dust ring widths, positions, and masses vary with increasing planet mass to identify ways in which observable ring characteristics could be used to constrain or estimate the masses of planets in protoplanetary discs. We run our simulations for three different aspect ratios to identify a connection between observable ring features and the pebble-isolation mass, which scales with the aspect ratio. We identify a number of observable ring properties and describe how these may be used to constrain planet masses. We find that:

\begin{itemize}
    \item The pebble-isolation mass corresponds to the relationship between the pressure gradients on either side of a pressure maximum (see Figure \ref{fig:dpdrs}) and on this basis, we propose that the observed gas surface density and pressure profiles of a disc can be used to identify whether a planet has a mass less than or greater than the pebble-isolation mass. Based on this, we present a new way of defining the pebble-isolation mass in terms of the gas pressure profile of a disc and demonstrate how this can be applied to the exoALMA sample \citep{Stadler2025}.
    \item Both the dust ring width and ring mass scale with planet mass for planets with masses up to the pebble-isolation mass. Planets more massive than this produce rings that are indistinguishable in terms of their widths or the mass they contain (see Figures \ref{fig:redges} and \ref{fig:rmasses}). 
    \item Planets below the pebble-isolation mass produce low mass, wide rings due to their dust traps allowing large amounts of solid material to escape. Conversely, higher mass planets produce compact, narrow rings.
    \item The inner edge location of a ring varies with Stokes number. We also confirm that it is linearly proportional to the planet's Hill radius, as previously shown by \citet{pinilla2012} and \citet{rosotti2016minimum}. The Stokes dependency should be taken into account if using the ring inner edge location (or gap outer edge location) to estimate planet masses (see Figure \ref{fig:redges}).
    \item The ring peak location is also linearly proportional to the planet's Hill radius but does not vary by Stokes number, so propose that it may serve as a better means for estimating the planet mass, if it can be accurately identified in observations (see Figure \ref{fig:rpeaks}). We prescribe an equation for calculating planet mass based on this (Equation \ref{eq:rpeak}). 
    \item In Section \ref{sec:pds70}, we apply our results to estimate the mass of PDS 70c and find good agreement with the findings of \citet{Hammond2025} and \citet{Ruzza2025}.
\end{itemize}

Observations of dust at all wavelengths can be used to make inferences about the masses of potential planets. Measurements of ring peak locations at any wavelength can be used to estimate the planet-to-star mass ratio. Ring widths measured in observations at known wavelengths can be used to estimate or constrain planet masses, depending on whether the planet mass falls above or below the pebble-isolation mass.\\

Most crucially, we show that the pebble-isolation mass can be defined in terms of the gas pressure gradients and therefore, gas observations can also be used to identify whether a planet's mass falls above or below the pebble-isolation mass of the system. The planet mass estimates obtained via the methods we describe can be compared with each other or with other proposed methods for estimating planet masses to validate the presence of a planet in a disc. For systems in which the disc's $\alpha$-viscosity parameter and aspect ratio at the planet's location are known or can be reasonably estimated, the methods detailed here provide a new way of calculating the masses of planets embedded in discs and furthering our understanding of how planets form in young discs. \\

\begin{acknowledgments}
The authors thank Lisa Patrascu, Gillian Bogert, and Joshua Adams for their work at McMaster during the early phases of this project.  We also thank the anonymous referee for a concise and useful report. AF acknowledges support from the Royal Society Enhancement Award. This research was enabled in part by support provided by McMaster University with computing support from a grant to REP from the Digital Research Alliance of Canada. 
J.S. acknowledges financial support from the Natural Sciences and Engineering Research Council of Canada (NSERC) through the Canada Graduate Scholarships Doctoral (CGS D) program, and from the Heising–Simons Foundation through the 51 Pegasi b Fellowship (grant \#2025-5885). REP is supported by a Discovery Grant from NSERC Canada. FM acknowledges support from The Royal Society Dorothy Hodgkin Fellowship. This paper makes use of the following ALMA data: ADS/JAO.ALMA\#2018.A.00030.S. ALMA is a partnership of ESO (representing its member states), NSF (USA) and NINS (Japan), together with NRC (Canada), NSTC and ASIAA (Taiwan), and KASI (Republic of Korea), in cooperation with the Republic of Chile. The Joint ALMA Observatory is operated by ESO, AUI/NRAO and NAOJ. The National Radio Astronomy Observatory and Green Bank Observatory are facilities of the U.S. National Science Foundation operated under cooperative agreement by Associated Universities, Inc.

\end{acknowledgments}

%



\software{FARGO3D \citep{masset2000fargo, llambay2016fargo}}



\appendix
\section{Ring edge locations}
\label{app1}

\begin{figure}
\centering
\includegraphics[width=\columnwidth]{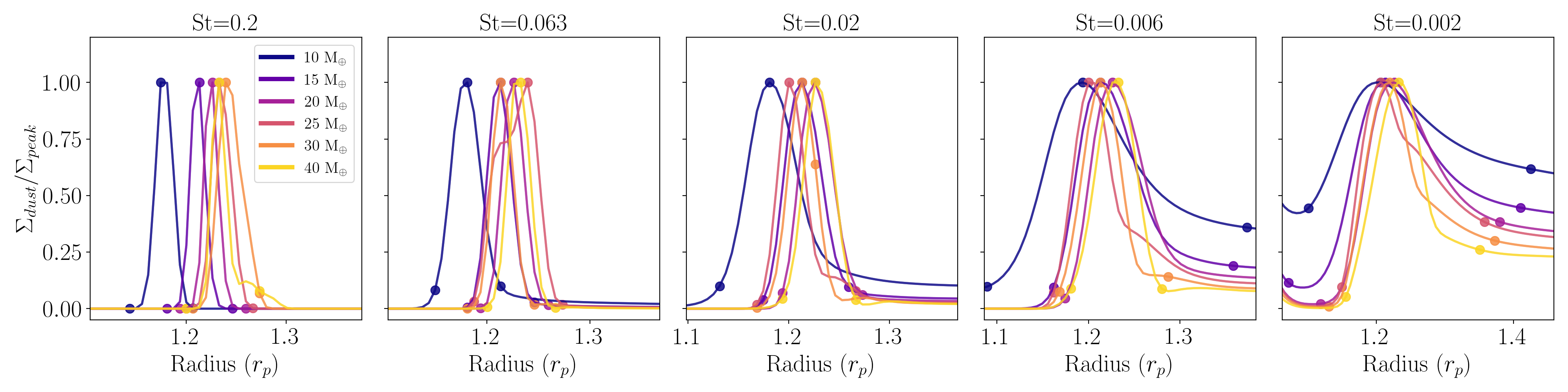}\par\vspace{1ex}
\includegraphics[width=\columnwidth]{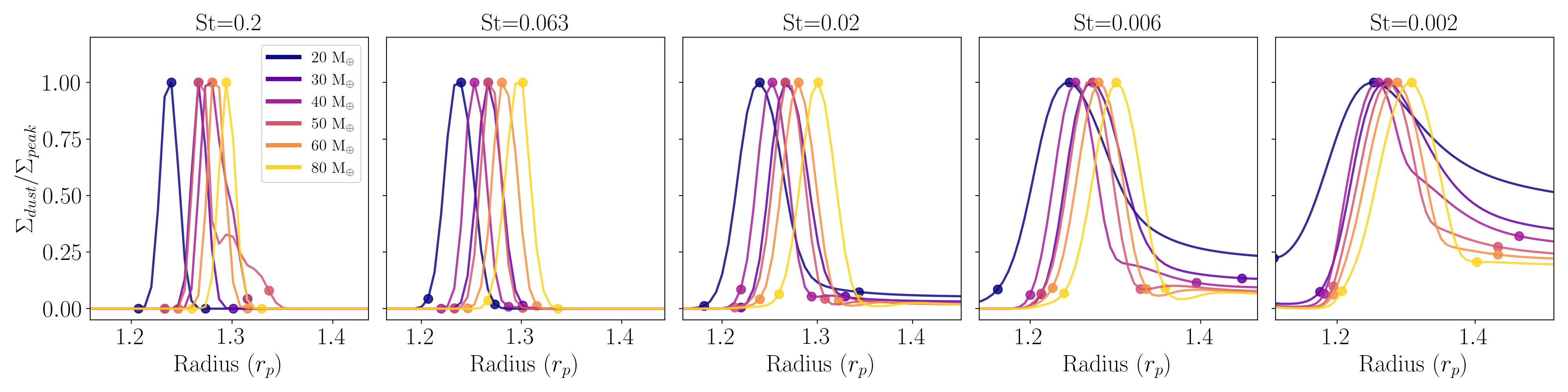}\par\vspace{1ex}
\includegraphics[width=\columnwidth]{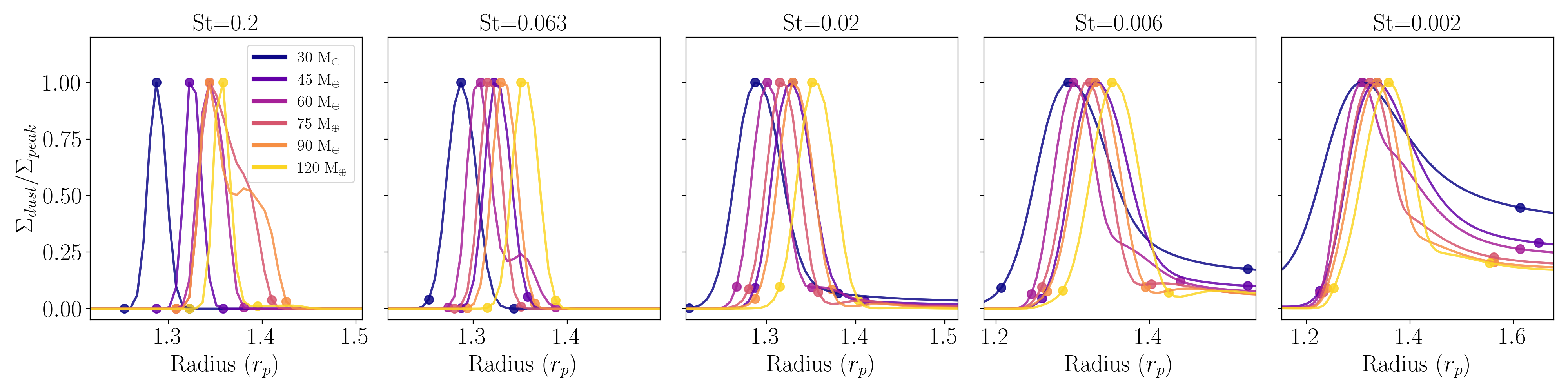}
\caption{Normalised dust density profiles and the ring edges identified in each case. Dots indicate the locations of the peaks and edges identifies for each profile. Top row: $h_{p}/r_{p}=0.05$, middle row: $h_{p}/r_{p}=0.06$, bottom row: $h_{p}/r_{p}=0.07$.}
\label{fig:1dsigmas}
\end{figure}

\begin{figure*}
    \centering
    \includegraphics[width=0.9\linewidth]{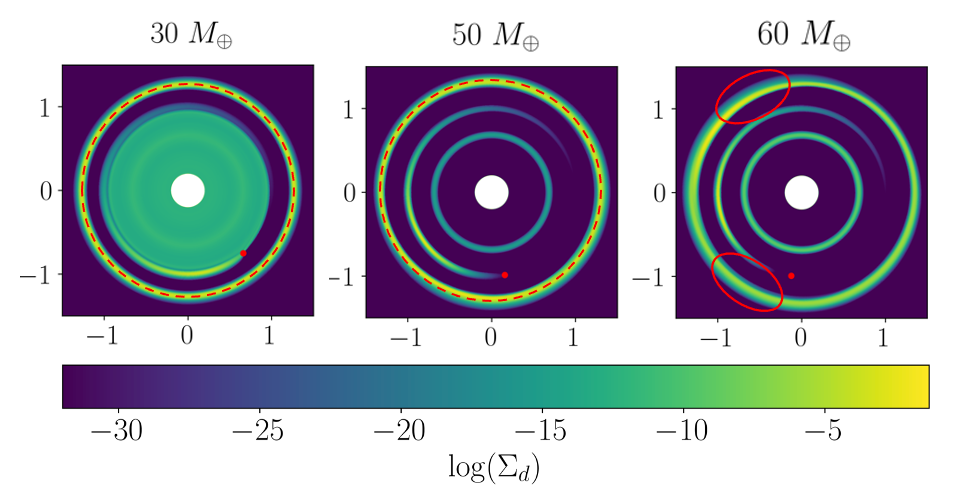}
    \caption{Snapshots of the 2D surface density profiles of St=0.2 dust for three different planet masses. All three snapshots are taken at 1500 orbits and are taken from simulations that used an aspect ratio of 0.06. The planets' locations are marked by red dots. The red dashed circles are centred on the origin and used to highlight azimuthal asymmetries in the left and middle plots. The red circles on the right plot highlight asymmetries which cancel out when azimuthlly averaged. }
    \label{fig:2dsigmas}
\end{figure*}

In Figure \ref{fig:1dsigmas}, we show the normalised 1D azimuthally averaged surface density profiles of dust of all Stokes numbers. These data are used to identify the ring peaks and edge locations according to the methods described in Section \ref{sec:ringfitting}. The identified ring peaks and edges are marked by dots. These profiles allow us to verify the reliability of our ring peak and edge identification methods, as well as explaining the origin of the some of the anomalous data points seen in Figures \ref{fig:rwidths}, \ref{fig:redges}, and \ref{fig:rpeaks}. In many cases, these outliers are caused by the presence of ``shoulders'' or ''bumps'', which cause the ring peaks to be misidentified or shifted, as well as altering the measured ring widths, as noted in Section \ref{sec:results:rwidth}. These shoulders only appear for higher mass planets so may be caused by the Rossby Wave Instability \citep{lovelace1999}. \\

Figure \ref{fig:2dsigmas} shows the final 2D surface density profiles of three simulations  with different planet masses but the same aspect ratio of 0.06. For ease of comparison, we focus on the same Stokes number, St=0.2, here. We present these three examples of the ways this instability can affect the measured ring peaks, edges, and widths. The left panel ($30 M_\oplus$) shows an example of a perfectly azimuthally symmetric ring, in which no shoulders are present. The data points corresponding to this simulation on Figures \ref{fig:rwidths} and \ref{fig:rpeaks} follow the expected trend and the 1D St=0.2 dust surface density profile for this simulation shows no asymmetric features.  In the middle panel we show an example of a ring with a non-axisymmetric narrowing of the outer edge, at the approximate coordinates of (-1.1, -0.5), which produces a visible shoulder in the corresponding 1D surface density profile in Figure \ref{fig:1dsigmas}. If the ring outer edge was not narrowed in this region, that shoulder would instead be part of a wider, more Gaussian-like density profile, with a peak located slightly further out. Instead, the shoulder causes the ring peak to be measured further in than expected (see Figure \ref{fig:rpeaks}). The right panel shows the $60 M_\oplus$ simulation, in which non-axisymmetric features are also present. However, these features are not visible in the equivalent 1D surface density profile. At around (1.2,-0.8), the dust surface density is lower on the exterior ring edge, whereas at (-0.5, -1), the lower density region (or shoulder) is seen on the ring's inner edge. These asymmetries cancel out when averaged, explaining why this feature is not seen in the 1D dust surface density profiles. However, it does still alter the morphology of the ring and therefore, the measured location of the ring peak, producing an anomalous data point on Figure \ref{fig:rpeaks}. \\

\bibliography{bibliography}{}

@ARTICLE{benisty2021,
       author = {{Benisty}, Myriam and {Bae}, Jaehan and {Facchini}, Stefano and {Keppler}, Miriam and {Teague}, Richard and {Isella}, Andrea and {Kurtovic}, Nicolas T. and {P{\'e}rez}, Laura M. and {Sierra}, Anibal and {Andrews}, Sean M. and {Carpenter}, John and {Czekala}, Ian and {Dominik}, Carsten and {Henning}, Thomas and {Menard}, Francois and {Pinilla}, Paola and {Zurlo}, Alice},
        title = "{A Circumplanetary Disk around PDS70c}",
      journal = {\apjl},
         year = 2021,
        month = jul,
       volume = {916},
       number = {1},
          eid = {L2},
        pages = {L2},
          doi = {10.3847/2041-8213/ac0f83},
archivePrefix = {arXiv},
       eprint = {2108.07123},
 primaryClass = {astro-ph.EP},
       adsurl = {https://ui.adsabs.harvard.edu/abs/2021ApJ...916L...2B}
}

@ARTICLE{gaia2021,
       author = {{Gaia Collaboration} and {Brown}, A.~G.~A. and {Vallenari}, A. and {Prusti}, T. and {de Bruijne}, J.~H.~J. and {Babusiaux}, C. and {Biermann}, M. and {Creevey}, O.~L. and {Evans}, D.~W. and {Eyer}, L. and {Hutton}, A. and {Jansen}, F. and {Jordi}, C. and {Klioner}, S.~A. and {Lammers}, U. and {Lindegren}, L. and {Luri}, X. and {Mignard}, F. and {Panem}, C. and {Pourbaix}, D. and {Randich}, S. and {Sartoretti}, P. and {Soubiran}, C. and {Walton}, N.~A. and {Arenou}, F. and {Bailer-Jones}, C.~A.~L. and {Bastian}, U. and {Cropper}, M. and {Drimmel}, R. and {Katz}, D. and {Lattanzi}, M.~G. and {van Leeuwen}, F. and {Bakker}, J. and {Cacciari}, C. and {Casta{\~n}eda}, J. and {De Angeli}, F. and {Ducourant}, C. and {Fabricius}, C. and {Fouesneau}, M. and {Fr{\'e}mat}, Y. and {Guerra}, R. and {Guerrier}, A. and {Guiraud}, J. and {Jean-Antoine Piccolo}, A. and {Masana}, E. and {Messineo}, R. and {Mowlavi}, N. and {Nicolas}, C. and {Nienartowicz}, K. and {Pailler}, F. and {Panuzzo}, P. and {Riclet}, F. and {Roux}, W. and {Seabroke}, G.~M. and {Sordo}, R. and {Tanga}, P. and {Th{\'e}venin}, F. and {Gracia-Abril}, G. and {Portell}, J. and {Teyssier}, D. and {Altmann}, M. and {Andrae}, R. and {Bellas-Velidis}, I. and {Benson}, K. and {Berthier}, J. and {Blomme}, R. and {Brugaletta}, E. and {Burgess}, P.~W. and {Busso}, G. and {Carry}, B. and {Cellino}, A. and {Cheek}, N. and {Clementini}, G. and {Damerdji}, Y. and {Davidson}, M. and {Delchambre}, L. and {Dell'Oro}, A. and {Fern{\'a}ndez-Hern{\'a}ndez}, J. and {Galluccio}, L. and {Garc{\'\i}a-Lario}, P. and {Garcia-Reinaldos}, M. and {Gonz{\'a}lez-N{\'u}{\~n}ez}, J. and {Gosset}, E. and {Haigron}, R. and {Halbwachs}, J.-L. and {Hambly}, N.~C. and {Harrison}, D.~L. and {Hatzidimitriou}, D. and {Heiter}, U. and {Hern{\'a}ndez}, J. and {Hestroffer}, D. and {Hodgkin}, S.~T. and {Holl}, B. and {Jan{\ss}en}, K. and {Jevardat de Fombelle}, G. and {Jordan}, S. and {Krone-Martins}, A. and {Lanzafame}, A.~C. and {L{\"o}ffler}, W. and {Lorca}, A. and {Manteiga}, M. and {Marchal}, O. and {Marrese}, P.~M. and {Moitinho}, A. and {Mora}, A. and {Muinonen}, K. and {Osborne}, P. and {Pancino}, E. and {Pauwels}, T. and {Petit}, J.-M. and {Recio-Blanco}, A. and {Richards}, P.~J. and {Riello}, M. and {Rimoldini}, L. and {Robin}, A.~C. and {Roegiers}, T. and {Rybizki}, J. and {Sarro}, L.~M. and {Siopis}, C. and {Smith}, M. and {Sozzetti}, A. and {Ulla}, A. and {Utrilla}, E. and {van Leeuwen}, M. and {van Reeven}, W. and {Abbas}, U. and {Abreu Aramburu}, A. and {Accart}, S. and {Aerts}, C. and {Aguado}, J.~J. and {Ajaj}, M. and {Altavilla}, G. and {{\'A}lvarez}, M.~A. and {{\'A}lvarez Cid-Fuentes}, J. and {Alves}, J. and {Anderson}, R.~I. and {Anglada Varela}, E. and {Antoja}, T. and {Audard}, M. and {Baines}, D. and {Baker}, S.~G. and {Balaguer-N{\'u}{\~n}ez}, L. and {Balbinot}, E. and {Balog}, Z. and {Barache}, C. and {Barbato}, D. and {Barros}, M. and {Barstow}, M.~A. and {Bartolom{\'e}}, S. and {Bassilana}, J.-L. and {Bauchet}, N. and {Baudesson-Stella}, A. and {Becciani}, U. and {Bellazzini}, M. and {Bernet}, M. and {Bertone}, S. and {Bianchi}, L. and {Blanco-Cuaresma}, S. and {Boch}, T. and {Bombrun}, A. and {Bossini}, D. and {Bouquillon}, S. and {Bragaglia}, A. and {Bramante}, L. and {Breedt}, E. and {Bressan}, A. and {Brouillet}, N. and {Bucciarelli}, B. and {Burlacu}, A. and {Busonero}, D. and {Butkevich}, A.~G. and {Buzzi}, R. and {Caffau}, E. and {Cancelliere}, R. and {C{\'a}novas}, H. and {Cantat-Gaudin}, T. and {Carballo}, R. and {Carlucci}, T. and {Carnerero}, M.~I. and {Carrasco}, J.~M. and {Casamiquela}, L. and {Castellani}, M. and {Castro-Ginard}, A. and {Castro Sampol}, P. and {Chaoul}, L. and {Charlot}, P. and {Chemin}, L. and {Chiavassa}, A. and {Cioni}, M.-R.~L. and {Comoretto}, G. and {Cooper}, W.~J. and {Cornez}, T. and {Cowell}, S. and {Crifo}, F. and {Crosta}, M. and {Crowley}, C. and {Dafonte}, C. and {Dapergolas}, A. and {David}, M. and {David}, P.},
        title = "{Gaia Early Data Release 3. Summary of the contents and survey properties}",
      journal = {\aap},
         year = 2021,
        month = may,
       volume = {649},
          eid = {A1},
        pages = {A1},
          doi = {10.1051/0004-6361/202039657},
archivePrefix = {arXiv},
       eprint = {2012.01533},
 primaryClass = {astro-ph.GA},
       adsurl = {https://ui.adsabs.harvard.edu/abs/2021A&A...649A...1G}
}

@ARTICLE{rosotti2016minimum,
       author = {{Rosotti}, Giovanni P. and {Juhasz}, Attila and {Booth}, Richard A. and {Clarke}, Cathie J.},
        title = "{The minimum mass of detectable planets in protoplanetary discs and the derivation of planetary masses from high-resolution observations}",
      journal = {\mnras},
         year = 2016,
        month = jul,
       volume = {459},
       number = {3},
        pages = {2790-2805},
          doi = {10.1093/mnras/stw691},
archivePrefix = {arXiv},
       eprint = {1603.02141},
 primaryClass = {astro-ph.EP},
       adsurl = {https://ui.adsabs.harvard.edu/abs/2016MNRAS.459.2790R}
}

@ARTICLE{meru2019ring,
       author = {{Meru}, Farzana and {Rosotti}, Giovanni P. and {Booth}, Richard A. and {Nazari}, Pooneh and {Clarke}, Cathie J.},
        title = "{Is the ring inside or outside the planet?: the effect of planet migration on dust rings}",
      journal = {\mnras},
         year = 2019,
        month = jan,
       volume = {482},
       number = {3},
        pages = {3678-3695},
          doi = {10.1093/mnras/sty2847},
archivePrefix = {arXiv},
       eprint = {1810.06573},
 primaryClass = {astro-ph.EP},
       adsurl = {https://ui.adsabs.harvard.edu/abs/2019MNRAS.482.3678M}
}

@ARTICLE{llambay2016fargo,
       author = {{Ben{\'\i}tez-Llambay}, Pablo and {Masset}, Fr{\'e}d{\'e}ric S.},
        title = "{FARGO3D: A New GPU-oriented MHD Code}",
      journal = {\apjs},
         year = 2016,
        month = mar,
       volume = {223},
       number = {1},
          eid = {11},
        pages = {11},
          doi = {10.3847/0067-0049/223/1/11},
archivePrefix = {arXiv},
       eprint = {1602.02359},
 primaryClass = {astro-ph.IM},
       adsurl = {https://ui.adsabs.harvard.edu/abs/2016ApJS..223...11B}
}

@INPROCEEDINGS{masset2000fargo,
       author = {{Masset}, F.~S.},
        title = "{FARGO: A Fast Eulerian Transport Algorithm for Differentially Rotating Disks}",
    booktitle = {Disks, Planetesimals, and Planets},
         year = 2000,
       editor = {{Garz{\'o}n}, G. and {Eiroa}, C. and {de Winter}, D. and {Mahoney}, T.~J.},
       series = {Astronomical Society of the Pacific Conference Series},
       volume = {219},
        month = jan,
        pages = {75},
       adsurl = {https://ui.adsabs.harvard.edu/abs/2000ASPC..219...75M}
}

@ARTICLE{goldreich1980,
       author = {{Goldreich}, P. and {Tremaine}, S.},
        title = "{Disk-satellite interactions.}",
      journal = {\apj},
         year = 1980,
        month = oct,
       volume = {241},
        pages = {425-441},
          doi = {10.1086/158356},
       adsurl = {https://ui.adsabs.harvard.edu/abs/1980ApJ...241..425G}
}

@ARTICLE{alma2015,
       author = {{ALMA Partnership} and {Brogan}, C.~L. and {P{\'e}rez}, L.~M. and {Hunter}, T.~R. and {Dent}, W.~R.~F. and {Hales}, A.~S. and {Hills}, R.~E. and {Corder}, S. and {Fomalont}, E.~B. and {Vlahakis}, C. and {Asaki}, Y. and {Barkats}, D. and {Hirota}, A. and {Hodge}, J.~A. and {Impellizzeri}, C.~M.~V. and {Kneissl}, R. and {Liuzzo}, E. and {Lucas}, R. and {Marcelino}, N. and {Matsushita}, S. and {Nakanishi}, K. and {Phillips}, N. and {Richards}, A.~M.~S. and {Toledo}, I. and {Aladro}, R. and {Broguiere}, D. and {Cortes}, J.~R. and {Cortes}, P.~C. and {Espada}, D. and {Galarza}, F. and {Garcia-Appadoo}, D. and {Guzman-Ramirez}, L. and {Humphreys}, E.~M. and {Jung}, T. and {Kameno}, S. and {Laing}, R.~A. and {Leon}, S. and {Marconi}, G. and {Mignano}, A. and {Nikolic}, B. and {Nyman}, L. -A. and {Radiszcz}, M. and {Remijan}, A. and {Rod{\'o}n}, J.~A. and {Sawada}, T. and {Takahashi}, S. and {Tilanus}, R.~P.~J. and {Vila Vilaro}, B. and {Watson}, L.~C. and {Wiklind}, T. and {Akiyama}, E. and {Chapillon}, E. and {de Gregorio-Monsalvo}, I. and {Di Francesco}, J. and {Gueth}, F. and {Kawamura}, A. and {Lee}, C. -F. and {Nguyen Luong}, Q. and {Mangum}, J. and {Pietu}, V. and {Sanhueza}, P. and {Saigo}, K. and {Takakuwa}, S. and {Ubach}, C. and {van Kempen}, T. and {Wootten}, A. and {Castro-Carrizo}, A. and {Francke}, H. and {Gallardo}, J. and {Garcia}, J. and {Gonzalez}, S. and {Hill}, T. and {Kaminski}, T. and {Kurono}, Y. and {Liu}, H. -Y. and {Lopez}, C. and {Morales}, F. and {Plarre}, K. and {Schieven}, G. and {Testi}, L. and {Videla}, L. and {Villard}, E. and {Andreani}, P. and {Hibbard}, J.~E. and {Tatematsu}, K.},
        title = "{The 2014 ALMA Long Baseline Campaign: First Results from High Angular Resolution Observations toward the HL Tau Region}",
      journal = {\apjl},
         year = 2015,
        month = jul,
       volume = {808},
       number = {1},
          eid = {L3},
        pages = {L3},
          doi = {10.1088/2041-8205/808/1/L3},
archivePrefix = {arXiv},
       eprint = {1503.02649},
 primaryClass = {astro-ph.SR},
       adsurl = {https://ui.adsabs.harvard.edu/abs/2015ApJ...808L...3A}
}

@ARTICLE{shakura1973,
       author = {{Shakura}, N.~I. and {Sunyaev}, R.~A.},
        title = "{Black holes in binary systems. Observational appearance.}",
      journal = {\aap},
         year = 1973,
        month = jan,
       volume = {24},
        pages = {337-355},
       adsurl = {https://ui.adsabs.harvard.edu/abs/1973A&A....24..337S}
}

@ARTICLE{bitsch2018,
       author = {{Bitsch}, Bertram and {Morbidelli}, Alessandro and {Johansen}, Anders and {Lega}, Elena and {Lambrechts}, Michiel and {Crida}, Aur{\'e}lien},
        title = "{Pebble-isolation mass: Scaling law and implications for the formation of super-Earths and gas giants}",
      journal = {\aap},
         year = 2018,
        month = apr,
       volume = {612},
          eid = {A30},
        pages = {A30},
          doi = {10.1051/0004-6361/201731931},
archivePrefix = {arXiv},
       eprint = {1801.02341},
 primaryClass = {astro-ph.EP},
       adsurl = {https://ui.adsabs.harvard.edu/abs/2018A&A...612A..30B}
}

@ARTICLE{lau2024,
       author = {{Lau}, Tommy Chi Ho and {Birnstiel}, Til and {Dr{\k{a}}{\.z}kowska}, Joanna and {Stammler}, Sebastian Markus},
        title = "{Sequential giant planet formation initiated by disc substructure}",
      journal = {arXiv e-prints},
         year = 2024,
        month = jun,
          eid = {arXiv:2406.12340},
        pages = {arXiv:2406.12340},
          doi = {10.48550/arXiv.2406.12340},
archivePrefix = {arXiv},
       eprint = {2406.12340},
 primaryClass = {astro-ph.EP},
       adsurl = {https://ui.adsabs.harvard.edu/abs/2024arXiv240612340L}
}

@INPROCEEDINGS{lin1993,
       author = {{Lin}, D.~N.~C. and {Papaloizou}, J.~C.~B.},
        title = "{On the Tidal Interaction Between Protostellar Disks and Companions}",
    booktitle = {Protostars and Planets III},
         year = 1993,
       editor = {{Levy}, Eugene H. and {Lunine}, Jonathan I.},
        month = jan,
        pages = {749},
       adsurl = {https://ui.adsabs.harvard.edu/abs/1993prpl.conf..749L}
}

@ARTICLE{morbidelli2012,
       author = {{Morbidelli}, A. and {Nesvorny}, D.},
        title = "{Dynamics of pebbles in the vicinity of a growing planetary embryo: hydro-dynamical simulations}",
      journal = {\aap},
         year = 2012,
        month = oct,
       volume = {546},
          eid = {A18},
        pages = {A18},
          doi = {10.1051/0004-6361/201219824},
archivePrefix = {arXiv},
       eprint = {1208.4687},
 primaryClass = {astro-ph.EP},
       adsurl = {https://ui.adsabs.harvard.edu/abs/2012A&A...546A..18M}
}

@ARTICLE{flock2015,
       author = {{Flock}, M. and {Ruge}, J.~P. and {Dzyurkevich}, N. and {Henning}, Th. and {Klahr}, H. and {Wolf}, S.},
        title = "{Gaps, rings, and non-axisymmetric structures in protoplanetary disks. From simulations to ALMA observations}",
      journal = {\aap},
         year = 2015,
        month = feb,
       volume = {574},
          eid = {A68},
        pages = {A68},
          doi = {10.1051/0004-6361/201424693},
archivePrefix = {arXiv},
       eprint = {1411.2736},
 primaryClass = {astro-ph.EP},
       adsurl = {https://ui.adsabs.harvard.edu/abs/2015A&A...574A..68F}
}

@ARTICLE{zhang2015,
       author = {{Zhang}, Ke and {Blake}, Geoffrey A. and {Bergin}, Edwin A.},
        title = "{Evidence of Fast Pebble Growth Near Condensation Fronts in the HL Tau Protoplanetary Disk}",
      journal = {\apjl},
         year = 2015,
        month = jun,
       volume = {806},
       number = {1},
          eid = {L7},
        pages = {L7},
          doi = {10.1088/2041-8205/806/1/L7},
archivePrefix = {arXiv},
       eprint = {1505.00882},
 primaryClass = {astro-ph.EP},
       adsurl = {https://ui.adsabs.harvard.edu/abs/2015ApJ...806L...7Z}
}

@ARTICLE{gonzalez2017,
       author = {{Gonzalez}, J. -F. and {Laibe}, G. and {Maddison}, S.~T.},
        title = "{Self-induced dust traps: overcoming planet formation barriers}",
      journal = {\mnras},
         year = 2017,
        month = may,
       volume = {467},
       number = {2},
        pages = {1984-1996},
          doi = {10.1093/mnras/stx016},
archivePrefix = {arXiv},
       eprint = {1701.01115},
 primaryClass = {astro-ph.EP},
       adsurl = {https://ui.adsabs.harvard.edu/abs/2017MNRAS.467.1984G}
}

@ARTICLE{gonzalez2015b,
       author = {{Gonzalez}, J. -F. and {Laibe}, G. and {Maddison}, S.~T. and {Pinte}, C. and {M{\'e}nard}, F.},
        title = "{ALMA images of discs: are all gaps carved by planets?}",
      journal = {\mnras},
         year = 2015,
        month = nov,
       volume = {454},
       number = {1},
        pages = {L36-L40},
          doi = {10.1093/mnrasl/slv120},
archivePrefix = {arXiv},
       eprint = {1509.00691},
 primaryClass = {astro-ph.EP},
       adsurl = {https://ui.adsabs.harvard.edu/abs/2015MNRAS.454L..36G}
}

@INPROCEEDINGS{paardekooper2004,
       author = {{Paardekooper}, S.~J. and {Mellema}, G.},
        title = "{Planets in Disks: A New Method for Hydrodynamic Disk Simulations}",
    booktitle = {Extrasolar Planets: Today and Tomorrow},
         year = 2004,
       editor = {{Beaulieu}, J. and {Lecavelier Des Etangs}, A. and {Terquem}, C.},
       series = {Astronomical Society of the Pacific Conference Series},
       volume = {321},
        month = dec,
        pages = {347},
       adsurl = {https://ui.adsabs.harvard.edu/abs/2004ASPC..321..347P}
}

@ARTICLE{youdin2005,
       author = {{Youdin}, Andrew N. and {Goodman}, Jeremy},
        title = "{Streaming Instabilities in Protoplanetary Disks}",
      journal = {\apj},
         year = 2005,
        month = feb,
       volume = {620},
       number = {1},
        pages = {459-469},
          doi = {10.1086/426895},
archivePrefix = {arXiv},
       eprint = {astro-ph/0409263},
 primaryClass = {astro-ph},
       adsurl = {https://ui.adsabs.harvard.edu/abs/2005ApJ...620..459Y}
}

@ARTICLE{bohlin1978ism,
       author = {{Bohlin}, R.~C. and {Savage}, B.~D. and {Drake}, J.~F.},
        title = "{A survey of interstellar H I from Lalpha absorption measurements. II.}",
      journal = {\apj},
         year = 1978,
        month = aug,
       volume = {224},
        pages = {132-142},
          doi = {10.1086/156357},
          url = {https://ui.adsabs.harvard.edu/abs/1978ApJ...224..132B},
       adsurl = {https://ui.adsabs.harvard.edu/abs/1978ApJ...224..132B}
}

@ARTICLE{dullemond2018,
       author = {{Dullemond}, Cornelis P. and {Birnstiel}, Tilman and {Huang}, Jane and {Kurtovic}, Nicol{\'a}s T. and {Andrews}, Sean M. and {Guzm{\'a}n}, Viviana V. and {P{\'e}rez}, Laura M. and {Isella}, Andrea and {Zhu}, Zhaohuan and {Benisty}, Myriam and {Wilner}, David J. and {Bai}, Xue-Ning and {Carpenter}, John M. and {Zhang}, Shangjia and {Ricci}, Luca},
        title = "{The Disk Substructures at High Angular Resolution Project (DSHARP). VI. Dust Trapping in Thin-ringed Protoplanetary Disks}",
      journal = {\apjl},
         year = 2018,
        month = dec,
       volume = {869},
       number = {2},
          eid = {L46},
        pages = {L46},
          doi = {10.3847/2041-8213/aaf742},
archivePrefix = {arXiv},
       eprint = {1812.04044},
 primaryClass = {astro-ph.EP},
       adsurl = {https://ui.adsabs.harvard.edu/abs/2018ApJ...869L..46D}
}

@ARTICLE{huang2025,
       author = {{Huang}, Pinghui and {Yu}, Fangyuan and {Lee}, Eve J. and {Dong}, Ruobing and {Bai}, Xue-Ning},
        title = "{Leaky Dust Traps in Planet-Embedded Protoplanetary Disks}",
      journal = {arXiv e-prints},
         year = 2025,
        month = mar,
          eid = {arXiv:2503.19026},
        pages = {arXiv:2503.19026},
          doi = {10.48550/arXiv.2503.19026},
archivePrefix = {arXiv},
       eprint = {2503.19026},
 primaryClass = {astro-ph.EP},
       adsurl = {https://ui.adsabs.harvard.edu/abs/2025arXiv250319026H}
}

@ARTICLE{zhang2018,
       author = {{Zhang}, Shangjia and {Zhu}, Zhaohuan and {Huang}, Jane and {Guzm{\'a}n}, Viviana V. and {Andrews}, Sean M. and {Birnstiel}, Tilman and {Dullemond}, Cornelis P. and {Carpenter}, John M. and {Isella}, Andrea and {P{\'e}rez}, Laura M. and {Benisty}, Myriam and {Wilner}, David J. and {Baruteau}, Cl{\'e}ment and {Bai}, Xue-Ning and {Ricci}, Luca},
        title = "{The Disk Substructures at High Angular Resolution Project (DSHARP). VII. The Planet-Disk Interactions Interpretation}",
      journal = {\apjl},
         year = 2018,
        month = dec,
       volume = {869},
       number = {2},
          eid = {L47},
        pages = {L47},
          doi = {10.3847/2041-8213/aaf744},
archivePrefix = {arXiv},
       eprint = {1812.04045},
 primaryClass = {astro-ph.EP},
       adsurl = {https://ui.adsabs.harvard.edu/abs/2018ApJ...869L..47Z}
}

@ARTICLE{ataiee2018,
       author = {{Ataiee}, S. and {Baruteau}, C. and {Alibert}, Y. and {Benz}, W.},
        title = "{How much does turbulence change the pebble isolation mass for planet formation?}",
      journal = {\aap},
         year = 2018,
        month = jul,
       volume = {615},
          eid = {A110},
        pages = {A110},
          doi = {10.1051/0004-6361/201732026},
archivePrefix = {arXiv},
       eprint = {1804.00924},
 primaryClass = {astro-ph.EP},
       adsurl = {https://ui.adsabs.harvard.edu/abs/2018A&A...615A.110A}
}

@ARTICLE{lodato2019,
       author = {{Lodato}, Giuseppe and {Dipierro}, Giovanni and {Ragusa}, Enrico and {Long}, Feng and {Herczeg}, Gregory J. and {Pascucci}, Ilaria and {Pinilla}, Paola and {Manara}, Carlo F. and {Tazzari}, Marco and {Liu}, Yao and {Mulders}, Gijs D. and {Harsono}, Daniel and {Boehler}, Yann and {M{\'e}nard}, Fran{\c{c}}ois and {Johnstone}, Doug and {Salyk}, Colette and {van der Plas}, Gerrit and {Cabrit}, Sylvie and {Edwards}, Suzan and {Fischer}, William J. and {Hendler}, Nathan and {Nisini}, Brunella and {Rigliaco}, Elisabetta and {Avenhaus}, Henning and {Banzatti}, Andrea and {Gully-Santiago}, Michael},
        title = "{The newborn planet population emerging from ring-like structures in discs}",
      journal = {\mnras},
         year = 2019,
        month = jun,
       volume = {486},
       number = {1},
        pages = {453-461},
          doi = {10.1093/mnras/stz913},
archivePrefix = {arXiv},
       eprint = {1903.05117},
 primaryClass = {astro-ph.SR},
       adsurl = {https://ui.adsabs.harvard.edu/abs/2019MNRAS.486..453L}
}

@ARTICLE{lambrechts2014,
       author = {{Lambrechts}, M. and {Johansen}, A. and {Morbidelli}, A.},
        title = "{Separating gas-giant and ice-giant planets by halting pebble accretion}",
      journal = {\aap},
         year = 2014,
        month = dec,
       volume = {572},
          eid = {A35},
        pages = {A35},
          doi = {10.1051/0004-6361/201423814},
archivePrefix = {arXiv},
       eprint = {1408.6087},
 primaryClass = {astro-ph.EP},
       adsurl = {https://ui.adsabs.harvard.edu/abs/2014A&A...572A..35L}
}

@ARTICLE{kanagawa2016,
       author = {{Kanagawa}, Kazuhiro D. and {Muto}, Takayuki and {Tanaka}, Hidekazu and {Tanigawa}, Takayuki and {Takeuchi}, Taku and {Tsukagoshi}, Takashi and {Momose}, Munetake},
        title = "{Mass constraint for a planet in a protoplanetary disk from the gap width}",
      journal = {\pasj},
         year = 2016,
        month = jun,
       volume = {68},
       number = {3},
          eid = {43},
        pages = {43},
          doi = {10.1093/pasj/psw037},
archivePrefix = {arXiv},
       eprint = {1603.03853},
 primaryClass = {astro-ph.EP},
       adsurl = {https://ui.adsabs.harvard.edu/abs/2016PASJ...68...43K}
}

@ARTICLE{Mathis1977,
       author = {{Mathis}, J.~S. and {Rumpl}, W. and {Nordsieck}, K.~H.},
        title = "{The size distribution of interstellar grains.}",
      journal = {\apj},
         year = 1977,
        month = oct,
       volume = {217},
        pages = {425-433},
          doi = {10.1086/155591},
       adsurl = {https://ui.adsabs.harvard.edu/abs/1977ApJ...217..425M}
}

@ARTICLE{Riols_Lesur2019,
       author = {{Riols}, A. and {Lesur}, G.},
        title = "{Spontaneous ring formation in wind-emitting accretion discs}",
      journal = {\aap},
         year = 2019,
        month = may,
       volume = {625},
          eid = {A108},
        pages = {A108},
          doi = {10.1051/0004-6361/201834813},
archivePrefix = {arXiv},
       eprint = {1904.07910},
 primaryClass = {astro-ph.EP},
       adsurl = {https://ui.adsabs.harvard.edu/abs/2019A&A...625A.108R}
}

@ARTICLE{Li_Youdin2021,
       author = {{Li}, Rixin and {Youdin}, Andrew N.},
        title = "{Thresholds for Particle Clumping by the Streaming Instability}",
      journal = {\apj},
         year = 2021,
        month = oct,
       volume = {919},
       number = {2},
          eid = {107},
        pages = {107},
          doi = {10.3847/1538-4357/ac0e9f},
archivePrefix = {arXiv},
       eprint = {2105.06042},
 primaryClass = {astro-ph.EP},
       adsurl = {https://ui.adsabs.harvard.edu/abs/2021ApJ...919..107L}
}

@ARTICLE{Keppler2018,
       author = {{Keppler}, M. and {Benisty}, M. and {M{\"u}ller}, A. and {Henning}, Th. and {van Boekel}, R. and {Cantalloube}, F. and {Ginski}, C. and {van Holstein}, R.~G. and {Maire}, A. -L. and {Pohl}, A. and {Samland}, M. and {Avenhaus}, H. and {Baudino}, J. -L. and {Boccaletti}, A. and {de Boer}, J. and {Bonnefoy}, M. and {Chauvin}, G. and {Desidera}, S. and {Langlois}, M. and {Lazzoni}, C. and {Marleau}, G. -D. and {Mordasini}, C. and {Pawellek}, N. and {Stolker}, T. and {Vigan}, A. and {Zurlo}, A. and {Birnstiel}, T. and {Brandner}, W. and {Feldt}, M. and {Flock}, M. and {Girard}, J. and {Gratton}, R. and {Hagelberg}, J. and {Isella}, A. and {Janson}, M. and {Juhasz}, A. and {Kemmer}, J. and {Kral}, Q. and {Lagrange}, A. -M. and {Launhardt}, R. and {Matter}, A. and {M{\'e}nard}, F. and {Milli}, J. and {Molli{\`e}re}, P. and {Olofsson}, J. and {P{\'e}rez}, L. and {Pinilla}, P. and {Pinte}, C. and {Quanz}, S.~P. and {Schmidt}, T. and {Udry}, S. and {Wahhaj}, Z. and {Williams}, J.~P. and {Buenzli}, E. and {Cudel}, M. and {Dominik}, C. and {Galicher}, R. and {Kasper}, M. and {Lannier}, J. and {Mesa}, D. and {Mouillet}, D. and {Peretti}, S. and {Perrot}, C. and {Salter}, G. and {Sissa}, E. and {Wildi}, F. and {Abe}, L. and {Antichi}, J. and {Augereau}, J. -C. and {Baruffolo}, A. and {Baudoz}, P. and {Bazzon}, A. and {Beuzit}, J. -L. and {Blanchard}, P. and {Brems}, S.~S. and {Buey}, T. and {De Caprio}, V. and {Carbillet}, M. and {Carle}, M. and {Cascone}, E. and {Cheetham}, A. and {Claudi}, R. and {Costille}, A. and {Delboulb{\'e}}, A. and {Dohlen}, K. and {Fantinel}, D. and {Feautrier}, P. and {Fusco}, T. and {Giro}, E. and {Gluck}, L. and {Gry}, C. and {Hubin}, N. and {Hugot}, E. and {Jaquet}, M. and {Le Mignant}, D. and {Llored}, M. and {Madec}, F. and {Magnard}, Y. and {Martinez}, P. and {Maurel}, D. and {Meyer}, M. and {M{\"o}ller-Nilsson}, O. and {Moulin}, T. and {Mugnier}, L. and {Orign{\'e}}, A. and {Pavlov}, A. and {Perret}, D. and {Petit}, C. and {Pragt}, J. and {Puget}, P. and {Rabou}, P. and {Ramos}, J. and {Rigal}, F. and {Rochat}, S. and {Roelfsema}, R. and {Rousset}, G. and {Roux}, A. and {Salasnich}, B. and {Sauvage}, J. -F. and {Sevin}, A. and {Soenke}, C. and {Stadler}, E. and {Suarez}, M. and {Turatto}, M. and {Weber}, L.},
        title = "{Discovery of a planetary-mass companion within the gap of the transition disk around PDS 70}",
      journal = {\aap},
         year = 2018,
        month = sep,
       volume = {617},
          eid = {A44},
        pages = {A44},
          doi = {10.1051/0004-6361/201832957},
archivePrefix = {arXiv},
       eprint = {1806.11568},
 primaryClass = {astro-ph.EP},
       adsurl = {https://ui.adsabs.harvard.edu/abs/2018A&A...617A..44K}
}

@ARTICLE{Haffert2019,
       author = {{Haffert}, S.~Y. and {Bohn}, A.~J. and {de Boer}, J. and {Snellen}, I.~A.~G. and {Brinchmann}, J. and {Girard}, J.~H. and {Keller}, C.~U. and {Bacon}, R.},
        title = "{Two accreting protoplanets around the young star PDS 70}",
      journal = {Nature Astronomy},
         year = 2019,
        month = jun,
       volume = {3},
        pages = {749-754},
          doi = {10.1038/s41550-019-0780-5},
archivePrefix = {arXiv},
       eprint = {1906.01486},
 primaryClass = {astro-ph.EP},
       adsurl = {https://ui.adsabs.harvard.edu/abs/2019NatAs...3..749H}
}

@ARTICLE{kanagawa2018,
       author = {{Kanagawa}, Kazuhiro D. and {Muto}, Takayuki and {Okuzumi}, Satoshi and {Tanigawa}, Takayuki and {Taki}, Tetsuo and {Shibaike}, Yuhito},
        title = "{Impacts of Dust Feedback on a Dust Ring Induced by a Planet in a Protoplanetary Disk}",
      journal = {\apj},
         year = 2018,
        month = nov,
       volume = {868},
       number = {1},
          eid = {48},
        pages = {48},
          doi = {10.3847/1538-4357/aae837},
archivePrefix = {arXiv},
       eprint = {1810.05635},
 primaryClass = {astro-ph.EP},
       adsurl = {https://ui.adsabs.harvard.edu/abs/2018ApJ...868...48K}
}

@ARTICLE{Zhu2014,
       author = {{Zhu}, Zhaohuan and {Stone}, James M. and {Rafikov}, Roman R. and {Bai}, Xue-ning},
        title = "{Particle Concentration at Planet-induced Gap Edges and Vortices. I. Inviscid Three-dimensional Hydro Disks}",
      journal = {\apj},
         year = 2014,
        month = apr,
       volume = {785},
       number = {2},
          eid = {122},
        pages = {122},
          doi = {10.1088/0004-637X/785/2/122},
archivePrefix = {arXiv},
       eprint = {1308.0648},
 primaryClass = {astro-ph.EP},
       adsurl = {https://ui.adsabs.harvard.edu/abs/2014ApJ...785..122Z}
}

@ARTICLE{Dong2018,
       author = {{Dong}, Ruobing and {Li}, Shengtai and {Chiang}, Eugene and {Li}, Hui},
        title = "{Multiple Disk Gaps and Rings Generated by a Single Super-Earth. II. Spacings, Depths, and Number of Gaps, with Application to Real Systems}",
      journal = {\apj},
         year = 2018,
        month = oct,
       volume = {866},
       number = {2},
          eid = {110},
        pages = {110},
          doi = {10.3847/1538-4357/aadadd},
archivePrefix = {arXiv},
       eprint = {1808.06613},
 primaryClass = {astro-ph.EP},
       adsurl = {https://ui.adsabs.harvard.edu/abs/2018ApJ...866..110D}
}

@ARTICLE{Bae2016,
       author = {{Bae}, Jaehan and {Nelson}, Richard P. and {Hartmann}, Lee},
        title = "{The Spiral Wave Instability Induced by a Giant Planet. I. Particle Stirring in the Inner Regions of Protoplanetary Disks}",
      journal = {\apj},
         year = 2016,
        month = dec,
       volume = {833},
       number = {2},
          eid = {126},
        pages = {126},
          doi = {10.3847/1538-4357/833/2/126},
archivePrefix = {arXiv},
       eprint = {1610.08502},
 primaryClass = {astro-ph.EP},
       adsurl = {https://ui.adsabs.harvard.edu/abs/2016ApJ...833..126B}
}

@ARTICLE{schreiber2018,
       author = {{Schreiber}, Andreas and {Klahr}, Hubert},
        title = "{Azimuthal and Vertical Streaming Instability at High Dust-to-gas Ratios and on the Scales of Planetesimal Formation}",
      journal = {\apj},
         year = 2018,
        month = jul,
       volume = {861},
       number = {1},
          eid = {47},
        pages = {47},
          doi = {10.3847/1538-4357/aac3d4},
archivePrefix = {arXiv},
       eprint = {1805.04326},
 primaryClass = {astro-ph.EP},
       adsurl = {https://ui.adsabs.harvard.edu/abs/2018ApJ...861...47S}
}

@ARTICLE{Doi2024,
       author = {{Doi}, Kiyoaki and {Kataoka}, Akimasa and {Liu}, Hauyu Baobab and {Yoshida}, Tomohiro C. and {Benisty}, Myriam and {Dong}, Ruobing and {Yamato}, Yoshihide and {Hashimoto}, Jun},
        title = "{Asymmetric Dust Accumulation of the PDS 70 Disk Revealed by ALMA Band 3 Observations}",
      journal = {\apjl},
         year = 2024,
        month = oct,
       volume = {974},
       number = {2},
          eid = {L25},
        pages = {L25},
          doi = {10.3847/2041-8213/ad7f51},
archivePrefix = {arXiv},
       eprint = {2408.09216},
 primaryClass = {astro-ph.EP},
       adsurl = {https://ui.adsabs.harvard.edu/abs/2024ApJ...974L..25D}
}

@ARTICLE{Alibert2018,
       author = {{Alibert}, Yann and {Venturini}, Julia and {Helled}, Ravit and {Ataiee}, Sareh and {Burn}, Remo and {Senecal}, Luc and {Benz}, Willy and {Mayer}, Lucio and {Mordasini}, Christoph and {Quanz}, Sascha P. and {Sch{\"o}nb{\"a}chler}, Maria},
        title = "{The formation of Jupiter by hybrid pebble-planetesimal accretion}",
      journal = {Nature Astronomy},
         year = 2018,
        month = aug,
       volume = {2},
        pages = {873-877},
          doi = {10.1038/s41550-018-0557-2},
archivePrefix = {arXiv},
       eprint = {1809.05383},
 primaryClass = {astro-ph.EP},
       adsurl = {https://ui.adsabs.harvard.edu/abs/2018NatAs...2..873A}
}

@ARTICLE{morbidelli2015,
       author = {{Morbidelli}, A. and {Lambrechts}, M. and {Jacobson}, S. and {Bitsch}, B.},
        title = "{The great dichotomy of the Solar System: Small terrestrial embryos and massive giant planet cores}",
      journal = {\icarus},
         year = 2015,
        month = sep,
       volume = {258},
        pages = {418-429},
          doi = {10.1016/j.icarus.2015.06.003},
archivePrefix = {arXiv},
       eprint = {1506.01666},
 primaryClass = {astro-ph.EP},
       adsurl = {https://ui.adsabs.harvard.edu/abs/2015Icar..258..418M}
}

@ARTICLE{Fasano2025,
       author = {{Fasano}, D. and {Benisty}, M. and {Curone}, P. and {Facchini}, S. and {Zagaria}, F. and {Yoshida}, T.~C. and {Doi}, K. and {Sierra}, A. and {Andrews}, S. and {Bae}, J. and {Isella}, A. and {Kurtovic}, N. and {P{\'e}rez}, L.~M. and {Pinilla}, P. and {Rampinelli}, L. and {Teague}, R.},
        title = "{Inner disc and circumplanetary material in the PDS 70 system: Insights from multi-epoch, multi-frequency ALMA observations}",
      journal = {\aap},
         year = 2025,
        month = jul,
       volume = {699},
          eid = {A373},
        pages = {A373},
          doi = {10.1051/0004-6361/202554959},
archivePrefix = {arXiv},
       eprint = {2506.11709},
 primaryClass = {astro-ph.EP},
       adsurl = {https://ui.adsabs.harvard.edu/abs/2025A&A...699A.373F}
}

@ARTICLE{Hammond2025,
       author = {{Hammond}, Iain and {Christiaens}, Valentin and {Price}, Daniel J. and {Blakely}, Dori and {Trevascus}, David and {Bonse}, Markus J. and {Cantalloube}, Faustine and {Marleau}, Gabriel-Dominique and {Pinte}, Christophe and {Juillard}, Sandrine and {Samland}, Matthias and {Thompson}, William and {Wallace}, Alex},
        title = "{Keplerian motion of a compact source orbiting the inner disc of PDS 70: a third protoplanet in resonance with b and c?}",
      journal = {\mnras},
         year = 2025,
        month = may,
       volume = {539},
       number = {2},
        pages = {1613-1627},
          doi = {10.1093/mnras/staf586},
archivePrefix = {arXiv},
       eprint = {2504.11127},
 primaryClass = {astro-ph.EP},
       adsurl = {https://ui.adsabs.harvard.edu/abs/2025MNRAS.539.1613H}
}

@ARTICLE{Stadler2025,
       author = {{Stadler}, Jochen and {Benisty}, Myriam and {Winter}, Andrew J. and {Izquierdo}, Andr{\'e}s F. and {Longarini}, Cristiano and {Galloway-Sprietsma}, Maria and {Curone}, Pietro and {Andrews}, Sean M. and {Bae}, Jaehan and {Facchini}, Stefano and {Rosotti}, Giovanni and {Teague}, Richard and {Barraza-Alfaro}, Marcelo and {Cataldi}, Gianni and {Cuello}, Nicol{\'a}s and {Czekala}, Ian and {Fasano}, Daniele and {Flock}, Mario and {Fukagawa}, Misato and {Garg}, Himanshi and {Hall}, Cassandra and {Hammond}, Iain and {Hilder}, Thomas and {Huang}, Jane and {Ilee}, John D. and {Kanagawa}, Kazuhiro and {Lesur}, Geoffroy and {Lodato}, Giuseppe and {Loomis}, Ryan A. and {Menard}, Francois and {Orihara}, Ryuta and {Pinte}, Christophe and {Price}, Daniel J. and {Yen}, Hsi-Wei and {Wafflard-Fernandez}, Gaylor and {Wilner}, David J. and {W{\"o}lfer}, Lisa and {Yoshida}, Tomohiro C. and {Zawadzki}, Brianna},
        title = "{exoALMA. VI. Rotating under Pressure: Rotation Curves, Azimuthal Velocity Substructures, and Gas Pressure Variations}",
      journal = {\apjl},
         year = 2025,
        month = may,
       volume = {984},
       number = {1},
          eid = {L11},
        pages = {L11},
          doi = {10.3847/2041-8213/adb152},
archivePrefix = {arXiv},
       eprint = {2504.20036},
 primaryClass = {astro-ph.EP},
       adsurl = {https://ui.adsabs.harvard.edu/abs/2025ApJ...984L..11S}
}

@ARTICLE{exoalma2025,
       author = {{Curone}, Pietro and {Facchini}, Stefano and {Andrews}, Sean M. and {Testi}, Leonardo and {Benisty}, Myriam and {Czekala}, Ian and {Huang}, Jane and {Ilee}, John D. and {Isella}, Andrea and {Lodato}, Giuseppe and {Loomis}, Ryan A. and {Stadler}, Jochen and {Winter}, Andrew J. and {Bae}, Jaehan and {Barraza-Alfaro}, Marcelo and {Cataldi}, Gianni and {Cuello}, Nicol{\'a}s and {Fasano}, Daniele and {Flock}, Mario and {Fukagawa}, Misato and {Galloway-Sprietsma}, Maria and {Garg}, Himanshi and {Hall}, Cassandra and {Izquierdo}, Andr{\'e}s F. and {Kanagawa}, Kazuhiro and {Lesur}, Geoffroy and {Longarini}, Cristiano and {Menard}, Francois and {Orihara}, Ryuta and {Pinte}, Christophe and {Price}, Daniel J. and {Rosotti}, Giovanni and {Teague}, Richard and {Wafflard-Fernandez}, Gaylor and {Wilner}, David J. and {W{\"o}lfer}, Lisa and {Yen}, Hsi-Wei and {Yoshida}, Tomohiro C. and {Zawadzki}, Brianna},
        title = "{exoALMA. IV. Substructures, Asymmetries, and the Faint Outer Disk in Continuum Emission}",
      journal = {\apjl},
         year = 2025,
        month = may,
       volume = {984},
       number = {1},
          eid = {L9},
        pages = {L9},
          doi = {10.3847/2041-8213/adc438},
archivePrefix = {arXiv},
       eprint = {2504.18725},
 primaryClass = {astro-ph.EP},
       adsurl = {https://ui.adsabs.harvard.edu/abs/2025ApJ...984L...9C}
}

@INPROCEEDINGS{lesur2023,
       author = {{Lesur}, G. and {Flock}, M. and {Ercolano}, B. and {Lin}, M.-K. and {Yang}, C. and {Barranco}, J.~A. and {Benitez-Llambay}, P. and {Goodman}, J. and {Johansen}, A. and {Klahr}, H. and {Laibe}, G. and {Lyra}, W. and {Marcus}, P.~S. and {Nelson}, R.~P. and {Squire}, J. and {Simon}, J.~B. and {Turner}, N.~J. and {Umurhan}, O.~M. and {Youdin}, A.~N.},
        title = "{Hydro-, Magnetohydro-, and Dust-Gas Dynamics of Protoplanetary Disks}",
    booktitle = {Protostars and Planets VII},
         year = 2023,
       editor = {{Inutsuka}, S. and {Aikawa}, Y. and {Muto}, T. and {Tomida}, K. and {Tamura}, M.},
       series = {Astronomical Society of the Pacific Conference Series},
       volume = {534},
        month = jul,
        pages = {465},
          doi = {10.48550/arXiv.2203.09821},
archivePrefix = {arXiv},
       eprint = {2203.09821},
 primaryClass = {astro-ph.EP},
       adsurl = {https://ui.adsabs.harvard.edu/abs/2023ASPC..534..465L}
}

@ARTICLE{izquierdo2025,
       author = {{Izquierdo}, Andr{\'e}s F. and {Stadler}, Jochen and {Galloway-Sprietsma}, Maria and {Benisty}, Myriam and {Pinte}, Christophe and {Bae}, Jaehan and {Teague}, Richard and {Facchini}, Stefano and {W{\"o}lfer}, Lisa and {Longarini}, Cristiano and {Curone}, Pietro and {Andrews}, Sean M. and {Barraza-Alfaro}, Marcelo and {Cataldi}, Gianni and {Cuello}, Nicol{\'a}s and {Czekala}, Ian and {Fasano}, Daniele and {Flock}, Mario and {Fukagawa}, Misato and {Garg}, Himanshi and {Hall}, Cassandra and {Hammond}, Iain and {Hilder}, Thomas and {Huang}, Jane and {Ilee}, John D. and {Isella}, Andrea and {Kanagawa}, Kazuhiro and {Lesur}, Geoffroy and {Lodato}, Giuseppe and {Loomis}, Ryan A. and {Orihara}, Ryuta and {Price}, Daniel J. and {Rosotti}, Giovanni and {Testi}, Leonardo and {Yen}, Hsi-Wei and {Wafflard-Fernandez}, Gaylor and {Wilner}, David J. and {Winter}, Andrew J. and {Yoshida}, Tomohiro C. and {Zawadzki}, Brianna},
        title = "{exoALMA. III. Line-intensity Modeling and System Property Extraction from Protoplanetary Disks}",
      journal = {\apjl},
         year = 2025,
        month = may,
       volume = {984},
       number = {1},
          eid = {L8},
        pages = {L8},
          doi = {10.3847/2041-8213/adc439},
archivePrefix = {arXiv},
       eprint = {2504.19986},
 primaryClass = {astro-ph.EP},
       adsurl = {https://ui.adsabs.harvard.edu/abs/2025ApJ...984L...8I}
}

@incollection{armitage2019,
  title={Physical processes in protoplanetary disks},
  author={Armitage, Philip J},
  booktitle={From Protoplanetary Disks to Planet Formation: Saas-Fee Advanced Course 45. Swiss Society for Astrophysics and Astronomy},
  pages={1--150},
  year={2019},
  publisher={Springer}
}

@ARTICLE{huang2018,
       author = {{Huang}, Jane and {Andrews}, Sean M. and {Dullemond}, Cornelis P. and {Isella}, Andrea and {P{\'e}rez}, Laura M. and {Guzm{\'a}n}, Viviana V. and {{\"O}berg}, Karin I. and {Zhu}, Zhaohuan and {Zhang}, Shangjia and {Bai}, Xue-Ning and {Benisty}, Myriam and {Birnstiel}, Tilman and {Carpenter}, John M. and {Hughes}, A. Meredith and {Ricci}, Luca and {Weaver}, Erik and {Wilner}, David J.},
        title = "{The Disk Substructures at High Angular Resolution Project (DSHARP). II. Characteristics of Annular Substructures}",
      journal = {\apjl},
         year = 2018,
        month = dec,
       volume = {869},
       number = {2},
          eid = {L42},
        pages = {L42},
          doi = {10.3847/2041-8213/aaf740},
archivePrefix = {arXiv},
       eprint = {1812.04041},
 primaryClass = {astro-ph.EP},
       adsurl = {https://ui.adsabs.harvard.edu/abs/2018ApJ...869L..42H}
}

@INPROCEEDINGS{weidenschilling1977a,
       author = {{Weidenschilling}, S.~J.},
        title = "{The solar nebula pressure gradient and its effect on planetesimal motions.}",
    booktitle = {IAU Colloquium 39: Comets, Asteroids, Meteorites: Interrelations, Evolution and Origins},
         year = 1977,
       editor = {{Delsemme}, A.~H.},
        month = jan,
        pages = {541-544},
       adsurl = {https://ui.adsabs.harvard.edu/abs/1977cami.coll..541W}
}

@INPROCEEDINGS{pinte2018,
       author = {{Pinte}, Christophe},
        title = "{Kinematic Evidence for an Embedded Protoplanet in a Circumstellar Disc}",
    booktitle = {Take a Closer Look},
         year = 2018,
        month = nov,
          eid = {99},
        pages = {99},
          doi = {10.5281/zenodo.1488968},
       adsurl = {https://ui.adsabs.harvard.edu/abs/2018tcl..confE..99P}
}

@ARTICLE{teague2018,
       author = {{Teague}, Richard and {Bae}, Jaehan and {Birnstiel}, Tilman and {Bergin}, Edwin A.},
        title = "{Evidence for a Vertical Dependence on the Pressure Structure in AS 209}",
      journal = {\apj},
         year = 2018,
        month = dec,
       volume = {868},
       number = {2},
          eid = {113},
        pages = {113},
          doi = {10.3847/1538-4357/aae836},
archivePrefix = {arXiv},
       eprint = {1810.04961},
 primaryClass = {astro-ph.EP},
       adsurl = {https://ui.adsabs.harvard.edu/abs/2018ApJ...868..113T}
}

@ARTICLE{exoalmaV2025,
       author = {{Galloway-Sprietsma}, Maria and {Bae}, Jaehan and {Izquierdo}, Andr{\'e}s F. and {Stadler}, Jochen and {Longarini}, Cristiano and {Teague}, Richard and {Andrews}, Sean M. and {Winter}, Andrew J. and {Benisty}, Myriam and {Facchini}, Stefano and {Rosotti}, Giovanni and {Zawadzki}, Brianna and {Pinte}, Christophe and {Fasano}, Daniele and {Barraza-Alfaro}, Marcelo and {Cataldi}, Gianni and {Cuello}, Nicol{\'a}s and {Curone}, Pietro and {Czekala}, Ian and {Flock}, Mario and {Fukagawa}, Misato and {Gardner}, Charles H. and {Garg}, Himanshi and {Hall}, Cassandra and {Huang}, Jane and {Ilee}, John D. and {Kanagawa}, Kazuhiro and {Lesur}, Geoffroy and {Lodato}, Giuseppe and {Loomis}, Ryan A. and {Menard}, Francois and {Orihara}, Ryuta and {Price}, Daniel J. and {Wafflard-Fernandez}, Gaylor and {Wilner}, David J. and {W{\"o}lfer}, Lisa and {Yen}, Hsi-Wei and {Yoshida}, Tomohiro C.},
        title = "{exoALMA. V. Gaseous Emission Surfaces and Temperature Structures}",
      journal = {\apjl},
         year = 2025,
        month = may,
       volume = {984},
       number = {1},
          eid = {L10},
        pages = {L10},
          doi = {10.3847/2041-8213/adc437},
archivePrefix = {arXiv},
       eprint = {2504.19902},
 primaryClass = {astro-ph.EP},
       adsurl = {https://ui.adsabs.harvard.edu/abs/2025ApJ...984L..10G}
}

@ARTICLE{exoalmaXVI2025,
       author = {{Barraza-Alfaro}, Marcelo and {Flock}, Mario and {B{\'e}thune}, William and {Teague}, Richard and {Bae}, Jaehan and {Benisty}, Myriam and {Cataldi}, Gianni and {Curone}, Pietro and {Czekala}, Ian and {Facchini}, Stefano and {Fasano}, Daniele and {Fukagawa}, Misato and {Galloway-Sprietsma}, Maria and {Garg}, Himanshi and {Hall}, Cassandra and {Huang}, Jane and {Ilee}, John D. and {Izquierdo}, Andr{\'e}s F. and {Kanagawa}, Kazuhiro and {Koch}, Eric W. and {Lesur}, Geoffroy and {Longarini}, Cristiano and {Loomis}, Ryan A. and {Orihara}, Ryuta and {Pinte}, Christophe and {Price}, Daniel J. and {Rosotti}, Giovanni and {Stadler}, Jochen and {Wafflard-Fernandez}, Gaylor and {Winter}, Andrew J. and {W{\"o}lfer}, Lisa and {Yen}, Hsi-Wei and {Yoshida}, Tomohiro C. and {Zawadzki}, Brianna},
        title = "{exoALMA. XVI. Predicting Signatures of Large-scale Turbulence in Protoplanetary Disks}",
      journal = {\apjl},
         year = 2025,
        month = may,
       volume = {984},
       number = {1},
          eid = {L21},
        pages = {L21},
          doi = {10.3847/2041-8213/adc42d},
archivePrefix = {arXiv},
       eprint = {2504.19853},
 primaryClass = {astro-ph.EP},
       adsurl = {https://ui.adsabs.harvard.edu/abs/2025ApJ...984L..21B}
}

@ARTICLE{pinilla2012,
       author = {{Pinilla}, P. and {Benisty}, M. and {Birnstiel}, T.},
        title = "{Ring shaped dust accumulation in transition disks}",
      journal = {\aap},
         year = 2012,
        month = sep,
       volume = {545},
          eid = {A81},
        pages = {A81},
          doi = {10.1051/0004-6361/201219315},
archivePrefix = {arXiv},
       eprint = {1207.6485},
 primaryClass = {astro-ph.EP},
       adsurl = {https://ui.adsabs.harvard.edu/abs/2012A&A...545A..81P}
}

@ARTICLE{lovelace1999,
       author = {{Lovelace}, R.~V.~E. and {Li}, H. and {Colgate}, S.~A. and {Nelson}, A.~F.},
        title = "{Rossby Wave Instability of Keplerian Accretion Disks}",
      journal = {\apj},
         year = 1999,
        month = mar,
       volume = {513},
       number = {2},
        pages = {805-810},
          doi = {10.1086/306900},
archivePrefix = {arXiv},
       eprint = {astro-ph/9809321},
 primaryClass = {astro-ph},
       adsurl = {https://ui.adsabs.harvard.edu/abs/1999ApJ...513..805L}
}

@ARTICLE{Birnstiel2024,
       author = {{Birnstiel}, Tilman},
        title = "{Dust Growth and Evolution in Protoplanetary Disks}",
      journal = {\araa},
         year = 2024,
        month = sep,
       volume = {62},
       number = {1},
        pages = {157-202},
          doi = {10.1146/annurev-astro-071221-052705},
archivePrefix = {arXiv},
       eprint = {2312.13287},
 primaryClass = {astro-ph.EP},
       adsurl = {https://ui.adsabs.harvard.edu/abs/2024ARA&A..62..157B}
}

@ARTICLE{Ruzza2025,
       author = {{Ruzza}, A. and {Lodato}, G. and {Rosotti}, G.~P. and {Armitage}, P.~J.},
        title = "{DBNets2.0: Simulation-based inference for planet-induced dust substructures in protoplanetary discs}",
      journal = {\aap},
         year = 2025,
        month = aug,
       volume = {700},
          eid = {A190},
        pages = {A190},
          doi = {10.1051/0004-6361/202554401},
archivePrefix = {arXiv},
       eprint = {2506.11200},
 primaryClass = {astro-ph.EP},
       adsurl = {https://ui.adsabs.harvard.edu/abs/2025A&A...700A.190R}
}

@ARTICLE{Wang2020,
       author = {{Wang}, Jason J. and {Ginzburg}, Sivan and {Ren}, Bin and {Wallack}, Nicole and {Gao}, Peter and {Mawet}, Dimitri and {Bond}, Charlotte Z. and {Cetre}, Sylvain and {Wizinowich}, Peter and {De Rosa}, Robert J. and {Ruane}, Garreth and {Liu}, Michael C. and {Absil}, Olivier and {Alvarez}, Carlos and {Baranec}, Christoph and {Choquet}, {\'E}lodie and {Chun}, Mark and {Defr{\`e}re}, Denis and {Delorme}, Jacques-Robert and {Duch{\^e}ne}, Gaspard and {Forsberg}, Pontus and {Ghez}, Andrea and {Guyon}, Olivier and {Hall}, Donald N.~B. and {Huby}, Elsa and {Jolivet}, A{\"\i}ssa and {Jensen-Clem}, Rebecca and {Jovanovic}, Nemanja and {Karlsson}, Mikael and {Lilley}, Scott and {Matthews}, Keith and {M{\'e}nard}, Fran{\c{c}}ois and {Meshkat}, Tiffany and {Millar-Blanchaer}, Maxwell and {Ngo}, Henry and {Orban de Xivry}, Gilles and {Pinte}, Christophe and {Ragland}, Sam and {Serabyn}, Eugene and {Catal{\'a}n}, Ernesto Vargas and {Wang}, Ji and {Wetherell}, Ed and {Williams}, Jonathan P. and {Ygouf}, Marie and {Zuckerman}, Ben},
        title = "{Keck/NIRC2 L'-band Imaging of Jovian-mass Accreting Protoplanets around PDS 70}",
      journal = {\aj},
         year = 2020,
        month = jun,
       volume = {159},
       number = {6},
          eid = {263},
        pages = {263},
          doi = {10.3847/1538-3881/ab8aef},
archivePrefix = {arXiv},
       eprint = {2004.09597},
 primaryClass = {astro-ph.EP},
       adsurl = {https://ui.adsabs.harvard.edu/abs/2020AJ....159..263W}
}
\bibliographystyle{aasjournal}



\end{document}